# Title: Highly siderophile elements were stripped from Earth's mantle by iron sulfide segregation

David C. Rubie[1*], Vera Laurenz[1], Seth A. Jacobson[1,2], Alessandro Morbidelli[2], Herbert Palme[3], Antje K. Vogel[1], Daniel J Frost[1]

[1]Bayerisches Geoinstitut, Bayreuth, Germany.

[2]Observatoire de la Cote d'Azur, Nice, France.

[3]Forschungsinstitut und Naturmuseum Senckenberg, Frankfurt, Germany.

*Correspondence to: dave.rubie@uni-bayreuth.de

**Abstract**: Highly siderophile elements (HSEs) are strongly depleted in the bulk silicate Earth (BSE) but are present in near-chondritic relative abundances. The conventional explanation is that the HSEs were stripped from the mantle by the segregation of metal during core formation but were added back in near-chondritic proportions by late accretion, after core formation had ceased. Here we show that metal-silicate equilibration and segregation during Earth's core formation actually increased HSE mantle concentrations because HSE partition coefficients are relatively low at the high pressures of core formation within Earth. The pervasive exsolution and segregation of iron sulfide liquid from silicate liquid (the "Hadean matte") stripped magma oceans of HSEs during cooling and crystallization, before late accretion, and resulted in slightly suprachondritic palladium/iridium and ruthenium/iridium ratios.

The formation of Earth's metallic core resulted from the segregation of liquid iron from silicates during accretion. This process partially removed siderophile (metal-loving) elements from the mantle by transporting them into the core. Moderately siderophile elements (MSEs), such as Ni, Co, W, and Mo, are variably depleted in the bulk silicate Earth (BSE) as a consequence of metal-silicate equilibration, because of their differing metal-silicate partition coefficients (*1, 2*). In contrast, the highly-siderophile elements (HSEs; Re, Os, Ir, Ru, Rh, Pt, Pd and Au) are present in near-chondritic relative abundances even though their metal-silicate partition coefficients (measured over the pressure range 0-18 GPa) vary by orders of magnitude (*3*). This has led to the widely-accepted hypothesis that the HSEs were stripped from the mantle by metal-silicate segregation and that the present concentrations were added by the late accretion of chondritic material after core formation had ceased (*4, 5, 6*). The mass of late-accreted material, as estimated from HSE concentrations, has also been used to determine the age of the Moon (*6*).

Unlike simple geochemical models of core formation, which unrealistically treat core-mantle equilibration as a single event (*7*), we modeled core formation as a multistage process because metal was delivered to Earth by accreting bodies throughout its accretion history. We followed the approach of Rubie et al. (*8*) in which evolving MSE abundances in Earth's mantle and core were modeled by integrating the dynamics of planetary accretion with the chemistry of core-mantle differentiation. In this approach, Earth's growth history comes from state-of-the-art *N*-body accretion simulations based on the "Grand Tack" scenario (*9–12*), which start with a protoplanetary disk consisting of 80 to 220 roughly Mars-sized embryos and several thousand smaller planetesimals distributed initially over heliocentric distances of 0.7 to 10 astronomical units (AU). For the HSE modeling presented here, the results are not dependent on the choice of the Grand Tack scenario because planets grow through embryo-embryo and embryo-planetesimal collisions in all accretion scenarios. Each collision is an accretion event, which delivers mass and energy to the growing planets, resulting in melting, magma ocean formation and an episode of core formation. Embryos and most planetesimals are assumed to have undergone early core-mantle differentiation. Unlike previous core formation models, the metal of projectile cores equilibrates with only a fraction of the target's mantle which is determined from a hydrodynamic model (*12, 13*). The compositions of metal and silicate produced in each core formation event are determined by a mass balance and element partitioning approach (*8, 14*). Five parameters are fit by least squares minimization so that the composition of the model Earth's mantle matches that of the BSE (*8*). Metal-silicate equilibration pressures are fit assuming that they are a constant fraction (~0.7 refined by least squares) of the target's core-mantle boundary pressure at the time of each impact, which on average is consistent with calculations of impact-induced melting during Earth's accretion (*15*). A heliocentric oxidation gradient model defines the bulk compositions of all starting bodies and is defined by four of the five fitting parameters (*8*) (Fig. S1A).

We investigated the evolution of mantle Ir, Pt, Pd and Ru concentrations during Earth's accretion and differentiation using our accretion and core formation model (*8*). Metal-silicate partition coefficients for the HSEs decrease with increasing $P$ and $T$ (*3*). In addition, Laurenz et al. have shown experimentally that increasing the sulfur content of the metal has a similar effect (*12, 16*). It is therefore essential to include sulfur in the initial bulk compositions of accreting bodies. Sulfur is a volatile element with a low 50% condensation temperature of 655 K at $10^{-4}$ bar (*17*). We therefore assumed that concentrations of S increased systematically with decreasing temperature and increasing heliocentric distance (*18*). We assumed that, before giant planet migration, fully-oxidized bodies that formed beyond Jupiter and Saturn (>6AU) contain the full complement of S (corresponding to 5.35 wt% in a CI composition) with concentrations decreasing along a linear gradient towards the Sun (Fig. S1B). We adjusted the heliocentric distance at which the S concentration becomes zero to 0.8 AU in order to obtain the Earth's bulk sulfur content (0.64 wt%). Although there are potential problems with this simple concentration-distance model, our main results do not depend on it (*12*). We used a partitioning model to determine the distribution of S between metallic and silicate liquids during each metal-silicate equilibration event (*12, 19*). Our ultimate objective was to obtain 1.7 to 2.0 wt.% S in Earth's core (*20*) and 200 to 250 ppm in the mantle (*21*) at the end of accretion.

Using high-pressure metal-silicate partition coefficients (*3, 16*), we considered the effects of metal-silicate equilibration and segregation on the evolution of Pt, Ru, Pd and Ir concentrations by including the modeled S abundances in the metallic liquid (Table S1). The final HSE concentrations were high and variably fractionated because of differing partition coefficients (*3, 16*), with the result that relative abundances in the mantle were strongly non-chondritic (Fig. 1A). Pd and Pt concentrations start to become especially high after the model Earth accreted ~60% of its mass and increased to strongly exceed BSE values by the end of accretion. Contrary to the conclusions of previous studies, accreted metal in these growth models actually added HSEs to Earth's mantle rather than removing them. In the case of differentiated planetesimals that underwent early (e.g. <3 My) core-mantle differentiation as a result of heating caused by the rapid decay of $^{26}$Al (*22*), HSE partition coefficients were extremely high ($10^6$-$10^{11}$) (*3*) at the low *P-T* conditions of planetesimal differentiation (≤0.3 GPa and ≤1900 K). This means that the HSEs partitioned almost entirely into the metallic cores of planetesimals, and to a lesser extent into embryo cores, during differentiation. In contrast, at the high *P-T* conditions of metal-silicate equilibration after Earth had accreted ~60% of its mass, HSE partition coefficients are lower, by two to five orders of magnitude than under the conditions of planetesimal differentiation (*3*). This resulted in HSEs in impactor cores being transferred to Earth's mantle by metal-silicate equilibration so that mantle concentrations increase to exceed BSE values (Fig. 1A). High HSE abundances also resulted from the accretion of fully-oxidized bodies and from the oxidation of accreted metal (delivered as small planetesimal cores and as dispersed metal in undifferentiated bodies) by water in the magma ocean (*8, 12*).

The results of Fig. 1A are based on the assumption that 100% of accreted metal equilibrates with silicate liquid. If only a limited fraction of metal equilibrates because of incomplete emulsification (*23*), concentrations of all four HSEs become even higher (Fig. S2A); this is because higher equilibration pressures are then required to reproduce MSE concentrations of Earth's mantle which cause further reductions of the HSE metal-silicate partition coefficients.

The final calculated mantle sulfur content exceeds the BSE concentration by a factor of ~30 (Fig. 1B) and, in addition, the S content of the core is very low (0.36 wt%). The high mantle concentrations that developed were not removed by metal-silicate equilibration and segregation (Fig. 1B) (*12*). In order to achieve a low BSE concentration (<200 ppm prior to late accretion) and a core concentration of 1.7 to 2.0 wt%, exsolution and segregation of FeS liquid to the core is required – an event that has been termed the "Hadean matte" (*24, 25*).

Sulfide liquid exsolves from a magma ocean when a solubility saturation level, termed the sulfur concentration at sulfide saturation (*SCSS*), has been exceeded (26). We have experimentally determined the SCSS for peridotite liquid experimentally at 7 to 21 GPa and 2373 to 2673 K (*16*) thus enabling concentrations in S-saturated magma oceans to be estimated by extrapolation using:

$$\ln(SCSS) = 14.2(\pm 1.18) - \frac{11032(\pm 3119)}{T} - \frac{379(\pm 82)P}{T} \qquad (1)$$

where *SCSS* is in ppm, *T* is in K and *P* is in GPa. Average values along magma ocean adiabats are much higher than the concentrations shown in Fig. 1B (Fig. 2). However, *SCSS* decreases strongly with decreasing temperature and becomes much lower close to the peridotite melting curve (Fig. 2). Thus droplets of immiscible FeS liquid exsolve from the silicate melt structure in a deep S-bearing magma ocean as it cools towards crystallization temperatures. This effect will be enhanced as the melt fraction is reduced by crystallization; because of its high density, the exsolved FeS segregates by sinking to the core (*12*).

We consider two end-member scenarios when modeling sulfide segregation, depending on magma oceans being either short-lived (*27*) or long-lived (*28*): (i) Sulfide segregation took place in multiple stages and occurred after each giant impact, provided *SCSS* was exceeded close to the peridotite melting curve, and (ii) FeS segregation took place in a single stage and occured only after the final giant impact and before late accretion. The final results of these two scenarios are almost identical, although the evolutionary paths are different.

The strong $P$-$T$ and depth dependences of SCSS in a deep magma ocean (Fig. 2) require a simple modeling approach because of the challenges in determining the amount of FeS liquid that exsolves and equilibrates chemically in a deep convecting magma ocean (*29*). We thus defined empirically an effective pressure $P_{eq-S}$ that describes the SCSS (using Eq. 1) and equilibration pressure for the entire magma ocean, assuming that the corresponding temperature lies between the liquidus and solidus of peridotite (*8*):

$$P_{eq-S} = k_S \times P_{CMB} \qquad (2)$$

Here $P_{CMB}$ is the core-mantle boundary pressure at the time of each FeS exsolution/segregation event and $k_S$ is a constant so that $P_{eq-S}$ increases as Earth accretes. The amount of FeS that exsolves is the excess that is present above the SCSS value calculated using Eqs. 1 and 2.

Because HSEs dissolve in S-bearing silicate melts as HSE-S species, they will be fractionated into sulfide liquid that exsolves from a magma ocean (*30*). We modeled the effect of segregating FeS liquid on mantle HSE concentrations, using our experimental data on the partitioning of Pt, Pd, Ru and Ir between FeS and peridotite liquids obtained at 7 to 21 GPa and 2373 to 2673K (*16*, Table S2). We assumed that equilibration between sulfide and silicate liquid occurs simultaneously with sulfide exsolution, at pressure $P_{eq-S}$ (Eq. 2). Because of high convection velocities (several meters per second) combined with a time scale of at least 1000 years to cool to crystallization temperatures (*27, 29*), sulfur should be well mixed in the magma ocean before exsolution starts. We therefore assumed that droplets of FeS liquid exsolve pervasively and equilibrate with the entire mantle at pressure $P_{eq-S}$, in contrast to the metallic liquid that segregated earlier, which equilibrates with only a limited fraction of the mantle at the base of the magma ocean (*8*).

The value of the adjustable parameter $k_S$ (Eq. 2) determines the final mantle S and HSE concentrations before late accretion (Fig. 3). Optimal results are obtained with $k_S \approx 0.44$. This fit requires the accretion of a late veneer to increase HSE concentrations to BSE values (*12*). Late accretion is modelled by terminating the segregation of sulfide liquid, which ends as a result of magma ocean solidification, because FeS liquid cannot percolate efficiently through crystalline mantle (*12, 31*).

The models that include both sulfide segregation and late accretion predict Ir, Pt and S concentrations that are close to BSE values, irrespective of whether FeS segregation occurred in a single stage or multiple stages (Figs. 4 and S3). Furthermore, we reproduced suprachondritic Pd/Ir and Ru/Ir ratios for the BSE (*4, 32*) (Fig. 4D). Sulfide segregation is very efficient at depleting the mantle in Pt and Ir, but it is less efficient at depleting Pd and Ru because these elements are less chalcophile than Pt and Ir at high *P-T* (*16*). This contrasts with metal-silicate partitioning behavior, in which Pd and Pt are the least siderophile (*3*). However, a number of other explanations have been proposed for suprachondritic Pd/Ir and Ru/Ir ratios (*4, 33*) that cannot readily be dismissed, especially considering the uncertainties on our results (Fig. 4D).

Accretion of a late veneer depends on the mantle being largely crystalline, because otherwise sulfide segregation in a magma ocean would simply continue, and HSE concentrations would never increase to BSE levels (*12*). The mixing of HSEs into convecting crystalline mantle is expected to have been a slow process, perhaps consistent with a previously estimated mixing time of ~1.5 Gy (*34*), and also may not have been complete (*35*). The age of the Moon has recently been determined using correlations between the time of the final giant impact and the mass of late-accreted material, derived from a large number of accretion simulations (*6*). Our results indicate that this correlation provides only a lower limit on the age of the Moon because it actually dates Earth's final magma ocean crystallization.

It has been argued that Earth accreted from the same reservoir before and after core formation because of correlated isotopic signatures of Mo and Ru; this argument is based on the assumption that Mo was added to the mantle mainly before late accretion, whereas Ru was added only with the late veneer (*36*). Here we show that this assumption is not valid because Ru concentrations increase in the mantle at an early stage of Earth accretion and long before addition of the late veneer, especially in the case of single-stage sulfide segregation (Fig. 4B, C).

We conclude that the addition of sulfur to Earth occurred over the entire history of accretion (Fig. 4A), refuting the assumption that all sulfur was added during late accretion (*37*). Although our results are based on assumptions about the distribution of S in the early solar system (*12*), it is unlikely that a plausible distribution could be found that would change our conclusion. The addition of S throughout accretion affects core formation models that use the elements W and Mo because S strongly influences their partitioning behavior (*38*). Last, we speculate that a possible low-density layer at the top of Earth's liquid outer core (*39*) could be the result of late FeS enrichment due to sulfide segregation.

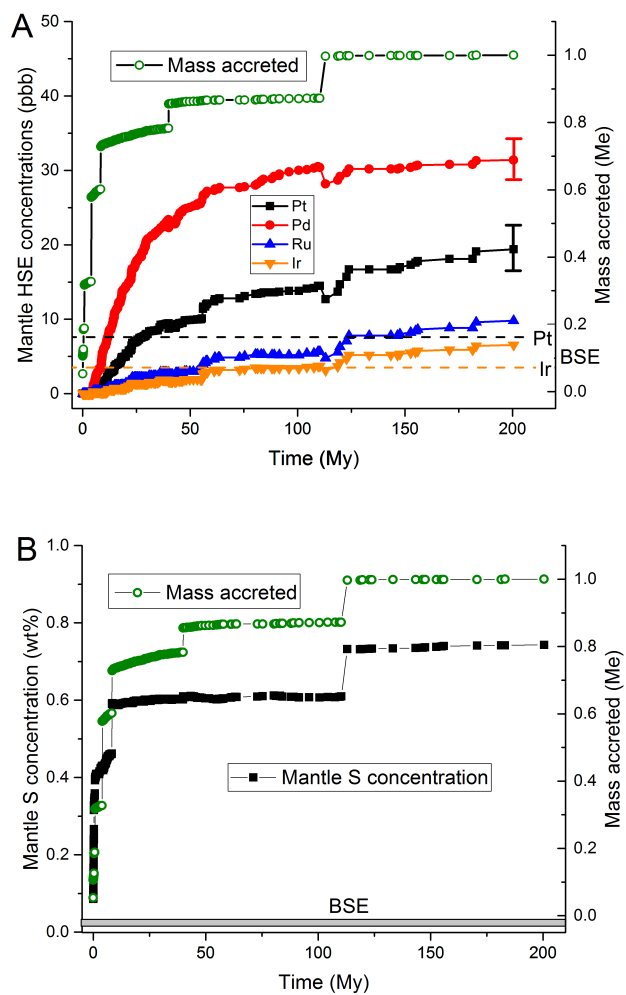

**Fig. 1. Evolution of HSEs and sulfur concentrations in the mantle during Earth's accretion based on metal-silicate equilibration and segregation.** Shown are mantle concentrations of (A) HSEs and (B) sulfur over time. Each symbol represents an impact, and "mass accreted" is the accumulated mass after each impact, normalized to Earth's current mass ($M_e$). The final giant impact, at 113 My, increases Earth's mass from $0.872 M_e$ to $0.997\ M_e$. BSE abundances (*21*) are shown by dashed lines in (A) and the gray bar in (B). Error bars, based on the propagation of uncertainties in the partitioning parameters, are shown for the final Pt and Pd concentrations (A); propagated uncertainties for Ru and Ir are ±0.6 and ±1.7 ppb, respectively. Results obtained

when only 50% of each batch of accreted metal equilibrates with silicate as a result of incomplete emulsification of impacting cores (*23*), are presented in Fig. S2. Although the oblique lines connecting the symbols show the general trends of mass and composition with time, they do not accurately represent the evolution paths, which in reality always involve a series of vertical steps.

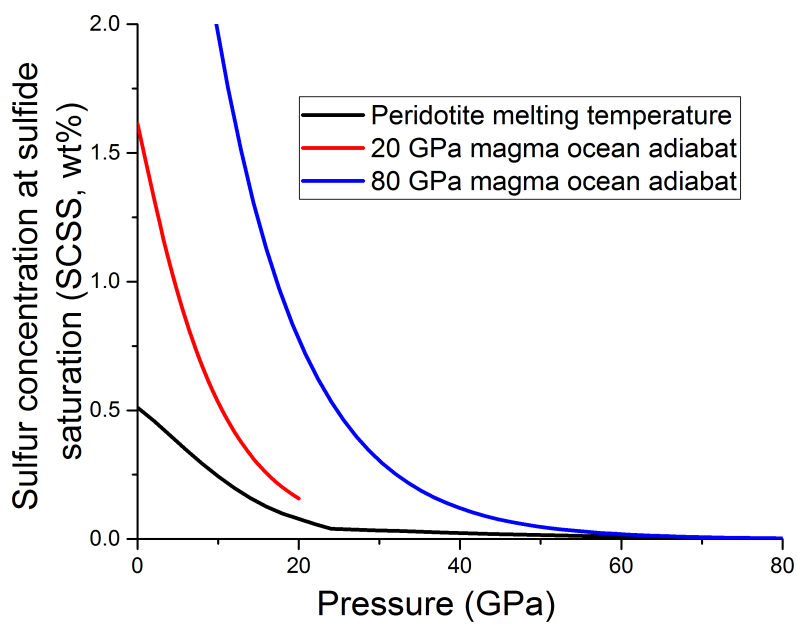

**Fig. 2**. **Sulfur concentrations at sulfide saturation (SCSS) in peridotite liquid, as a function of pressure.** Equation 1, from Laurenz et al. (*16*), has been used to calculate concentrations at temperatures about midway between the peridotite liquidus and solidus (*8*) and along adiabatic temperature profiles for magma oceans with basal pressures of 20 GPa and 80 GPa.

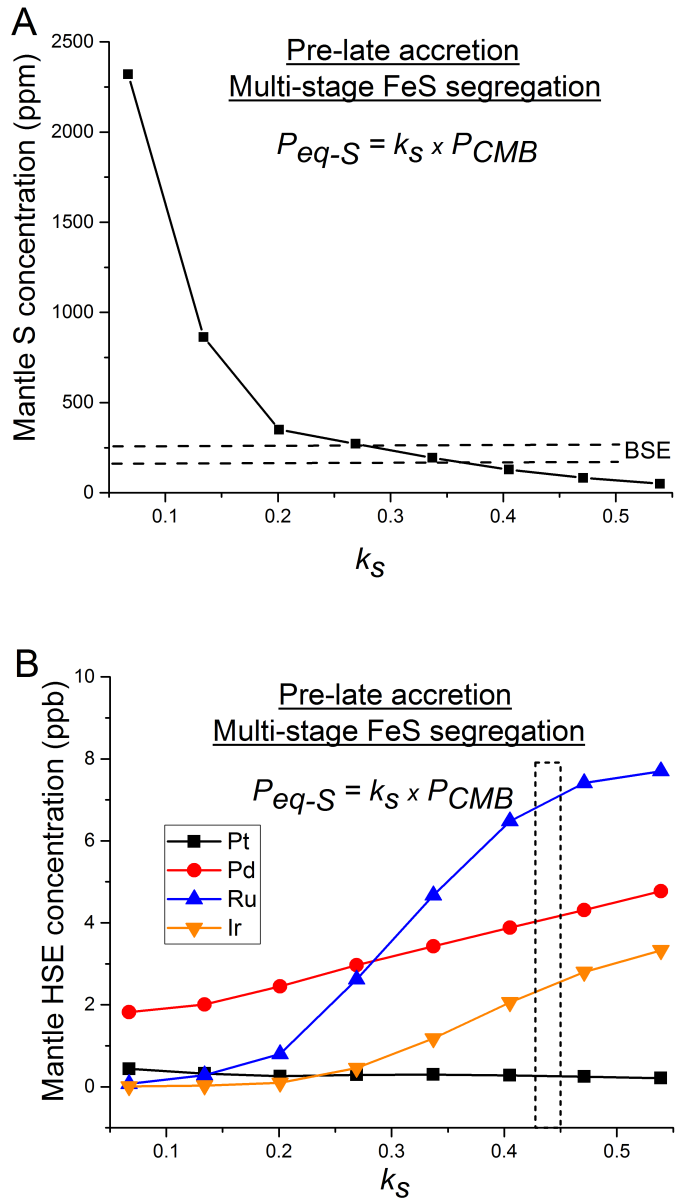

**Fig. 3. Final mantle concentrations after multi-stage sulfide segregation but before late accretion.** Shown are the final concentrations of (A) sulfur (with the BSE concentration shown by the horizontal dashed-outline bar) and (B) HSEs that result from multistage sulfide segregation, without late accretion, as a function of $k_s$ (Eq. 2). When late accretion is also modeled, BSE abundances are best reproduced with $k_s \approx 0.44$ [vertical dashed outline bar in (B)] (Fig. 4). The concentrations of Pd and Ru are elevated because these elements are the least chalcophile at high *P-T*. A similar result is obtained for single-stage sulfide segregation.

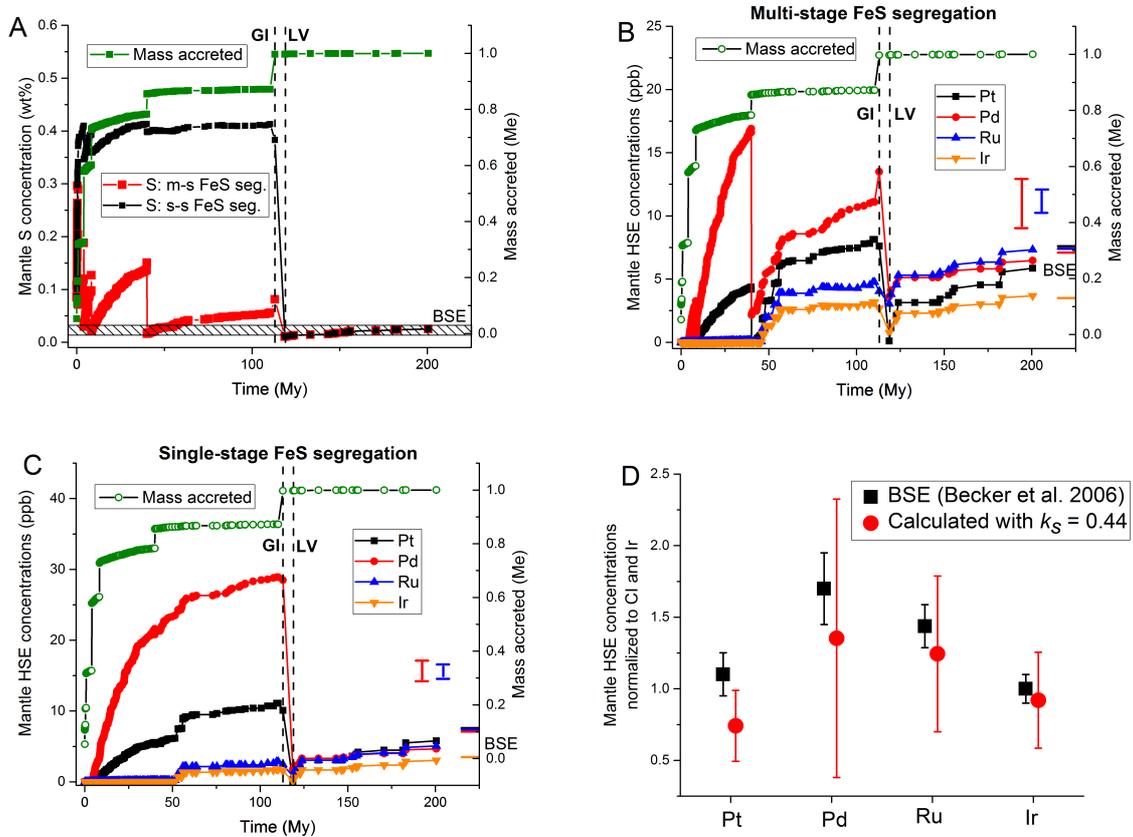

**Fig. 4**. **Final results based on metal-silicate segregation, sulfide segregation, and late accretion.** The evolution of (A) sulfur and (B and C) HSE concentrations with time are shown with $k_s$ = 0.44 (Eq. 2). The accretion history is shown in (A) to (C). Results are shown for multistage FeS segregation (m-s FeS seg.) and late single-stage segregation (s-s FeS seg.) at 118 My. Mass accreted is as in Fig. 1. The vertical dashed lines in (A) to (C) show the time of the final giant impact (GI) at 113 My and the start of late veneer accretion (LV) at 119 My. Horizontal bars show BSE concentrations. Error bars, based on the propagation of uncertainties in the partitioning parameters, are shown in (B) and (C) for the final Pd and Ru concentrations; propagated uncertainties for Pt and Ir are ±0.3 and ±0.6 ppb respectively. As in Fig. 1, the oblique lines connecting the symbols show the general trends of mass and composition with time, but they do not accurately represent the evolution paths. (D) Final calculated HSE values for multistage FeS segregation, normalized to Ir and CI chondrite composition, compared with BSE values (*32*). The error bars on the calculated values are based on propagating the partitioning parameter uncertainties (Tables S1 and S2).

**References and Notes:**

1. B. J. Wood, M. J. Walter, J. Wade, Accretion of the Earth and segregation of its core. *Nature* **441**, 825-833 (2006).
2. D. C. Rubie, F. Nimmo, H. J. Melosh, "Formation of the Earth's Core" in *Treatise on Geophysics,* D. J. Stevenson, Ed.*,* Vol. 9 *Evolution of the Earth,* Gerald Schubert, Ed., (Elsevier, Oxford, ed. 2, 2015) pp. 43-79.
3. U. Mann, D. J. Frost, D. C. Rubie, H. Becker, A. Audétat, Partitioning of Ru, Rh, Pd, Re, Ir and Pt between liquid metal and silicate at high pressures and high temperatures - Implications for the origin of highly siderophile element concentrations in the Earth's mantle. *Geochim. Cosmochim. Acta* **84**, 593–613 (2012).
4. R. J. Walker, K. Bermingham, J. Liu, I. S. Puchtel, M. Touboul, E. A. Worsham, In search of planetary building blocks. *Chem. Geol.* **411**, 125-142 (2015).
5. C.-L. Chou, Fractionation of siderophile elements in the Earth's upper mantle. *Proc. Lunar Planet Sci.* **9**, 219-230 (1978).
6. S. A. Jacobson, A. Morbidelli, S. N. Raymond, D. P. O'Brien, K. J. Walsh, D. C. Rubie, Highly siderophile elements in the Earth's mantle as a clock for the Moon-forming impact. *Nature* **508**, 84-87 (2014).
7. M. J. Walter, E. Cottrell, Assessing uncertainty in geochemical models for core formation in Earth. *Earth Planet Sci. Lett.* **365**, 165-176 (2013).
8. D. C. Rubie, S. A. Jacobson, A. Morbidelli, D. P. O'Brien, E. D. Young, J. de Vries, F. Nimmo, H. Palme, D. J. Frost, Accretion and differentiation of the terrestrial planets with implications for the compositions of early-formed Solar System bodies and accretion of water. *Icarus* **248**, 89-108 (2015).
9. D. P. O'Brien, K. J. Walsh, A. Morbidelli, S. N. Raymond, A. M. Mandell, Water delivery and giant impacts in the 'Grand Tack. *Icarus* **239**, 74-84 (2014).
10. S. A. Jacobson, A. Morbidelli, Lunar and terrestrial planet formation in the Grand Tack scenario. *Proc. Royal Soc*. **A 372**, 20130174 (2014).

11. K. J. Walsh, A. Morbidelli, S. N. Raymond, D. P. O'Brien, A. M. Mandell, A low mass for Mars from Jupiter's early gas-driven migration. *Nature* **475**, 206-209 (2011).

12. Materials are available as supplementary materials on Science Online.

13. R. Deguen, P. Olson, P. Cardin, Experiments on turbulent metal- silicate mixing in a magma ocean. *Earth Planet. Sci. Lett.*, **310**, 303-313 (2011).

14. D. C. Rubie, D. J. Frost, U. Mann, Y. Asahara, K. Tsuno, F. Nimmo, P. Kegler, A. Holzheid, H. Palme, Heterogeneous accretion, composition and core-mantle differentiation of the Earth. *Earth Planet. Sci. Lett.* **301**, 31-42 (2011).

15. J. de Vries, F. Nimmo, H. J. Melosh, S. A. Jacobson, A. Morbidelli, D. C. Rubie, Impact-induced melting during accretion of the Earth. *Prog. Earth Planet. Sci*. **3:7** (2016).

16. V. Laurenz, D. C. Rubie, D. J. Frost, A. K. Vogel, The importance of sulfur for the behaviour of highly-siderophile elements during Earth's differentiation. *Geochim. Cosmochim. Acta*, in press, doi:10.1016/j.gca.2016.08.012.

17. K. Lodders, Solar System abundances and condensation temperatures of the elements. *The Astrophys. J.* **591**, 1220-1247 (2003).

18. P. Cassen, Models for the fractionation of moderately volatile elements in the solar nebula. *MAPS* **31**, 793-806 (1996).

19. A. Boujibar, D. Andrault, A. Bouhifd, N. Bolfan-Casanova, J.-L. Devidal, N. Trcera, Metal-silicate partitioning of sulphur, new experimental and thermodynamic constraints on planetary accretion. *Earth Planet. Sci. Lett.* **391**, 42-54 (2014).

20. G. Dreibus, H. Palme, Cosmochemical constraints on the sulfur content in the Earth's core. *Geochim. Cosmochim. Acta* **60**, 1125–1130 (1996).

21. H. Palme, H. St. C. O'Neill, "Cosmochemical estimates of mantle composition" in *Treatise on Geochemistry,* H. D. Holland and K. K. Turekian, Eds., Vol. 3 *The Mantle and Core,* R. W. Carlson, Ed. (Elsevier-Pergamon, Oxford, ed. 2, 2014), pp. 1-39.

22. T. S. Kruijer, P. Sprung, T. Kleine, I. Leya, C. Burkhardt, R. Wieler, Hf-W chronometry of core formation in planetesimals inferred from weakly irradiated iron meteorites*. Geochim. Cosmochim. Acta* **99,** 287-304 (2012).

23. T. W. Dahl, D. J. Stevenson, Turbulent mixing of metal and silicate during planetary accretion – and interpretation of the Hf-W chronometer. *Earth Planet. Sci. Lett.* **295**, 177-186 (2010).

24. H. St. C. O'Neill, The origin of the Moon and the early history of the Earth – A chemical model. Part 2: The Earth. *Geochim. Cosmochim. Acta* **55**, 1159-1172 (1991).

25. P. S. Savage, F. Moynier, H. Chen, G. Shofner, J. Siebert, J. Badro, I. S. Puchtel, Copper isotope evidence for large-scale sulphide fractionation during Earth's differentiation. *Geochem. Persp. Let.* **1**, 53-64 (2015).

26. J. A. Mavrogenes, H. St. C. O'Neill, The relative effects of pressure, temperature and oxygen fugacity on the solubility of sulfide in mafic magmas. *Geochim. Cosmochim. Acta* **63**, 1173-1180 (1999).

27. V. S. Solomatov, "Fluid dynamics of a terrestrial magma ocean" in *Origin of the Earth and Moon*, R. M. Canup, K. Righter, Eds. (Univ. Ariz. Press, 2000), pp. 323-338.

28. Y. Abe, Thermal and chemical evolution of the terrestrial magma ocean. *Phys. Earth Planet. Int*. **100**, 27-39 (1997).

29. D. C. Rubie, H. J. Melosh, J. E. Reid, C. Liebske, K. Righter, Mechanisms of metal-silicate equilibration in the terrestrial magma ocean. *Earth Planet. Sci. Lett.* **205**, 239-255 (2003).

30. V. Laurenz, R. O. C. Fonesca, C. Ballhaus, K. P. Jochum, A. Heuser, P. J. Sylvester, The solubility of palladium and ruthenium in picritic melts: 2. The effect of sulfur. *Geochim. Cosmochim. Acta* **108**, 172-183 (2013).

31. H. Terasaki, D. J. Frost, D. C. Rubie, F. Langenhorst, Percolative core formation in planetesimals. *Earth Planet. Sci. Lett.* **273**, 132-137 (2008).

32. H. Becker, M. F. Horan, R. J. Walker, S. Gao, J.-P. Lorand, R. L. Rudnick, Highly siderophile element composition of the Earth's primitive upper mantle: Constraints from new data on peridotite massifs and xenoliths. *Geochim. Cosmochim. Acta* **70**, 4528-4550 (2006).

33. M. Fischer-Gödde, H. Becker, Osmium isotope and highly siderophile element constraints on ages and nature of meteoritic components in ancient lunar impact rocks. *Geochim. Cosmochim. Acta* **77**, 135-156 (2012).

34. W. D. Maier, S. J. Barnes, I. H. Campbell, M. L. Fiorentini, P. Peltonen, S.-J. Barnes, R. H. Smithies, Progressive mixing of meteoritic veneer into the early Earth's deep mantle. *Nature* **460**, 620-623 (2009).

35. H. Rizo, R. J. Walker, R. W. Carlson, M. F. Horan, S. Mukhopadhyay, V. Manthos, D. Francis, M. G. Jackson, Preservation of Earth-forming events in the tungsten isotopic composition of modern flood basalts. *Science* **352**, 809-812 (2016).

36. N. Dauphas, A. M. Davis, B. Marty, L. Reisberg, The cosmic molybdenum-ruthenium isotope correlation. *Earth Planet. Sci. Lett*. **226**, 465-475 (2004).

37. F. Albarede, Volatile accretion history of the terrestrial planets and dynamic implications. *Nature* **461**, 1227-1233 (2009).

38. J. Wade, B. J. Wood, J. Tuff, Metal-silicate partitioning of Mo and W at high pressures and temperatures: Evidence for late accretion of Sulphur to the Earth. *Geochim. Cosmochim. Acta* **85**, 58-74 (2012).

39. G. Helffrich, S. Kaneshima, Outer-core compositional stratification from observed core wave speed profiles. *Nature* **468**, 807-810 (2010).

40. F. Masset, M. Snellgrove, Reversing type II migration: resonance trapping of a lighter giant protoplanet. *Mon. Not. R. Astron. Soc*. **320**, L55–L59 (2001).

41. A. Morbidelli, A. Crida, The dynamics of Jupiter and Saturn in the gaseous protoplanetary disk. *Icarus* **191**, 158-171 (2007).

42. A. Pierens, R. P. Nelson, On the formation and migration of giant planets in circumbinary discs. *A&A* **483**, 633-642 (2008).

43. B. M. S. Hansen, Formation of the Terrestrial Planets from a Narrow Annulus. *ApJ* **703**, 1131-1140 (2009).

44. J. E. Chambers, Making more terrestrial planets. *Icarus* **152**, 205-224 (2001).

45. E. Kokubo, S. Ida, Oligarchic Growth of Protoplanets. *Icarus* **131**, 171-178 (1998).

46. E. Kokubo, J. Kominami, S. Ida, Formation of Terrestrial Planets from Protoplanets. I. Statistics of Basic Dynamical Properties. *Astrophys. J.* **642**, 1131-1139 (2006).

47. B. J. Wood, E. S. Kiseeva, F. J. Mirolo, Accretion and core formation: The effects of sulfur on metal-silicate partition coefficients. *Geochim. Cosmochim. Acta* **145**, 248-267 (2014).

48. J. D. Kendall, H. J. Melosh, Differentiated planetesimal impacts into a terrestrial magma ocean: Fate of the iron core. *Earth Planet. Sci. Lett, 448,* 24-33 (2016).

49. R. G. Kraus, S. Root, R. W. Lemke, S. T. Stewart, S. B. Jacobsen, T. R. Mattsson, Impact vaporization of planetesimal cores in the late stages of planet formation. *Nature Geosci.* **8**, 269-272 (2015).

50. R. Deguen, M. Landeau, P. Olson, Turbulent metal-silicate mixing, fragmentation, and equilibration in magma oceans. *Earth Planet. Sci. Lett*. **391**, 274-287 (2014).

51. H. Samuel, A re-evaluation of metal diapir breakup and equilibration in terrestrial magma oceans. *Earth Planet. Sci. Lett*. **313-314**, 105-114 (2012).
52. J. F. Rudge, T. Kleine, B. Bourdon, Broad bounds on Earth's accretion and core formation constrained by geochemical models. *Nature Geosci.* **3**, 439-443 (2010).
53. F. Nimmo, D. P. O'Brien, T. Kleine, Tungsten isotope evolution during late-stage accretion: Constraints on Earth-Moon equilibration. *Earth Planet. Sci. Lett*. **292**, 363-370 (2010).
54. J. Siebert, J. Badro, D. Antonangeli, F. J. Ryerson, Metal–silicate partitioning of Ni and Co in a deep magma ocean. *Earth Planet. Sci. Lett*. **321-322**, 189-197 (2012).
55. R. A. Fischer, Y. Nakajima, A. J. Campbell, D. J. Frost, D. Harries, F. Langenhorst, N. Miyajima, K. Pollock, D. C. Rubie, High pressure metal-silicate partitioning of Ni, Co, V, Cr, Si and O. *Geochim. Cosmochim. Acta* **167**, 177-194 (2015).
56. Japan Society for the Promotion of Science, *Steelmaking Data Sourcebook* (Gordon and Breach Science Publishers, New York, 1988) pp. 273–297.
57. K. Tsuno, D. J. Frost, D. C. Rubie, The effects of nickel and sulphur on the core-mantle partitioning of oxygen in Earth and Mars. *Phys. Earth Planet. Int*. **185**, 1-12 (2011).
58. C. Defouilloy, P. Cartigny, N. Assayag, F. Moynier, J.-A. Barrat, High-precision sulfur isotope composition of enstatite meteorites and implications of the formation and evolution of their parent bodies. *Geochim. Cosmochim. Acta* **172**, 393–409 (2016).
59. O. Nebel, K. Mezger, W. van Westrenen, Rubidium isotopes in primitive chondrites: Constraints on Earth's volatile element depletion and lead isotope evolution. *Earth Planet. Sci. Lett*. **305**, 309–316 (2011).
60. H. Terasaki, S. Urakawa, D. C. Rubie, K. Funakoshi, T. Sakamaki, Y. Shibazaki, S. Ozawa, E. Ohtani, Interfacial tension of Fe-Si liquid at high pressure: Implications for liquid Fe-alloy droplet size in magma oceans. *Phys. Earth Planet. Int*. **202-203**, 1-6 (2012).
61. A. Holzheid, Sulphide melt distribution in partially molten silicate aggregates: implications to core formation scenarios in terrestrial planets. *Eur. J. Mineral.* **25**, 267-277 (2013).
62. Cerantola, V., Walte, N., Rubie, D.C. (2015) Deformation of a crystalline system with two immiscible liquids: Implications for early core-mantle differentiation. Earth and Planetary Science Letters 417, 67-77.

**Acknowledgments:**

D.C.R., V.L., S.A.J., A.M. and H.P. were supported by the European Research Council Advanced Grant "ACCRETE" (contract number 290568) and A.K.V. was supported by the German Science Foundation (DFG) Priority Programme SPP1385 "The First 10 Million Years of the Solar System – a Planetary Materials Approach (grant Ru1323/2). We thank H.J. Melosh and R.J. Walker for discussions and three reviewers for their helpful and constructive comments. Data files with final results (Fig. 4) are available in the Supplementary Materials.

**Supplementary Materials:**
Supplementary text
Figures S1-S4
Tables S1-S2
Databases S1 to S2.

# Supplementary Materials for

# Highly siderophile elements were stripped from Earth's mantle by iron sulfide segregation

David C. Rubie, Vera Laurenz, Seth A. Jacobson, Alessandro Morbidelli, Herbert Palme, Antje K. Vogel, Daniel J Frost

correspondence to: dave.rubie@uni-bayreuth.de

**This PDF file includes:**

Supplementary Text
Figs. S1 to S4
Tables S1 to S2
Captions for databases S1 to S2

## 1. Supplementary Text

Late veneer and late accretion

These terms have been used with slightly different meanings by different authors – e.g. Walker et al. (*4*) and Jacobson et al. (*6*). Here we use "late veneer" to mean all material that was accreted to Earth after the end of core formation and sulfide segregation. We use the term "late accretion" to describe the physical process by which the late veneer was added.

Accretion/core formation model

The accretion/core formation model is identical to one of six published models (*8*) but with the addition of two volatile elements (Na and S) and four highly-siderophile elements (HSEs: Pt, Ru, Pd and Ir).

Grand Tack accretion scenario

The Grand tack scenario is the first accretion model that couples the processes of migration of the giant planets and formation of terrestrial planets. Originally proposed by Walsh et al. (*11*), this scenario is based on previous results of hydro-dynamical simulations showing that a Jupiter-mass planet alone in the protoplanetary disk would migrate inwards but the couple Jupiter-Saturn, once trapped in resonance, would migrate outwards (*40, 41, 42*). The Grand Tack scenario thus postulates that Jupiter formed first, at a distance of 3-4 AU from the Sun, and migrated inwards. Saturn, after reaching a mass close to its present one, also started to migrate towards the Sun and caught Jupiter in resonance. At this point, Jupiter and Saturn reversed the direction of migration and started to move towards the outer disk. The reversal of migration is dubbed the "tack", hence the name of the model. The embryo and planetesimal populations in the region (~3-6 AU) swept by the inwards-then-outwards migration of Jupiter are strongly depleted (Fig. S1A). The Grand Tack scenario postulates that the tack occurred when Jupiter was at about 1.5 AU from the Sun, so that the resulting depletion of mass outside of 1 AU can explain the final small mass of Mars (*43*), which could not be reproduced by earlier classical accretion models.

From the perspective of terrestrial planet formation, the Grand Tack model predicts a wide mixing of embryos and planetesimals that originally accreted in different parts of the disk. The local material, indigenous of the region ~1 AU, is mixed with material originally from the 1-3 AU region and pushed inwards by the resonances with Jupiter during the inward migration phase of the giant planet. This process of dynamical mixing also boosts the initial accretion rate of the terrestrial planets relative to the classical scenario (*44*), because it breaks the mutual dynamical isolation of the planetary embryos resulting from their growth in oligarchic fashion (*45, 46*). Finally, when the outer giant planets migrate outwards, multiple primitive and water-rich planetesimals from the giant planet zone are scattered towards the terrestrial planet region, thus contributing significantly to Earth's water budget (*9*).

The Grand Tack scenario is also successful in explaining the orbital distribution of the terrestrial planets (*10*), such as their low angular momentum deficit (which is a measure of the small deviations of the terrestrial planet orbits from perfect co-planar circles). Provided some assumptions hold true on the mass ratio between embryos and planetesimals, it also predicts that the growth of the Earth, although initially very rapid, can be protracted for ~100 My; in this case the amount of material accreted after the last giant impact is small (less than 1% of the Earth-mass), consistent with the amount of "late veneer" inferred from mantle HSE abundances (*5, 6, 33*).

From the compositional point of view, given reasonable initial chemical gradients in the disk, as outlined below, the Grand Tack scenario is consistent with the chemistry of the major non-volatile lithophile and weakly-siderophile elements in the mantle and the oxidation state of the Earth (*8*).

Structure and chemistry of the protoplanetary disk

The Grand Tack simulation selected for this study is 4:1-0.5-8 because it produced a model Earth close to 1 AU with a final mass close to one Earth mass ($M_e$). In addition, the final giant impact occurs at 113 My and the mass of material accreted following this impact is about $0.003 M_e$. In the simulation number "4:1-0.5-8", "4:1" indicates the ratio of the total masses in the embryo and planetesimal populations respectively, and "0.5" signifies that the initial embryo mass is 0.5 × mass of Mars. Based on these parameters, ten simulations were run with very slight variations in the initial orbital characteristics of the starting bodies: the last term ("8") is the run number within the set of 10 simulations. The protoplanetary disk consists of 87 embryos, distributed between 0.7-3.0 AU, and 2836 planetesimals, each of mass $3.9 \times 10^{-4} M_e$, initially distributed between 0.7-3.0 AU and 6.0-9.5 AU (the region between 3.0 and 6.0 AU is cleared of planetesimals by the accretion of Jupiter and Saturn).

The bulk chemistry of each of the embryos and planetesimals is specified <u>prior to the migration of Jupiter and Saturn</u> by assuming Solar System bulk composition ratios (CI chondritic) for all non-volatile elements (Mg, Si, Ni, Co, Cr and Pd) but with refractory element concentrations (Ca, Al, Nb, Ta, Pt, Ru and Ir) that are enhanced by 22% (11% for V) relative to CI (normalized to Mg), as justified previously (*14*). Oxygen and water contents are two critical compositional variables. The composition of a model Earth's mantle, calculated from our accretion/core formation model, is fit to that of the bulk silicate Earth (BSE). The exploration of a broad parameter space results in the optimal composition-distance model for primitive bodies shown in Fig. S1A (*8*). Here compositions close to the Sun (<1.1 AU) are highly reduced with 99.9% of Fe being present as metal and with ~18% of total Si initially dissolved in the metal. At heliocentric distances >1.1 AU, compositions become increasing oxidized and beyond 6.9 AU they are fully oxidized and contain 20 wt% $H_2O$. Four least-squares fitting parameters are indicated in Fig. S1A as red arrows and consist of two compositional parameters and two distance parameters. The distance $\delta(3)$ was fixed to be 6.9 AU by requiring a final mantle $H_2O$ content of ~1000 ppm.

Using a similar approach, we include the volatile element S (and also for comparison the lithophile/volatile element Na) in the current model by imposing compositional gradients on starting bodies (Fig. S1B). These gradients reflect high volatility at low heliocentric distances because of high temperatures and an increased tendency to condense as temperatures decrease with increasing heliocentric distance in the nebula (*18*). The Na gradient is adjusted in order to obtain a final mantle (BSE) concentration of 2600 ppm (*21*) assuming that Na does not partition into the core. In the case of S, the aim is to obtain the bulk Earth's S content which results when bodies at ≤0.8 AU contain zero

S. The steeper gradient for S compared to Na is qualitatively consistent with the lower 50% condensation temperatures of S at $10^{-4}$ bar of 655 K compared to 953 K for Na. (*17*).

The sulfur gradient of Fig. S1B is likely to be an oversimplification of reality. For example, it implies that highly reduced compositions from ≤1 AU should contain no S – whereas enstatite chondrites, which are highly reduced, contain up to 5-6 wt% S. Such discrepancies may possibly be a consequence of the evolution of protoplanetary disk chemistry with time. We show below that provided there is sufficient S to provide Earth with its bulk S content (0.64 wt%), our main conclusions are independent of its distribution in the protoplanetary disk.

Core formation model

We modelled the evolution of some moderately siderophile non-volatile element abundances in Earth's mantle and core by integrating the dynamical process of planetary accretion with the chemistry of core-mantle differentiation (*8*). Planets grow through collisions with embryos and planetesimals which (apart from fully-oxidized C-type bodies) are assumed to be differentiated into core and mantle prior to accretion. Each impact delivers mass and energy to a growing planet, with the energy resulting in melting, magma ocean formation and an episode of core formation. The initial bulk compositions of all bodies are defined in terms of oxygen and $H_2O$ contents, as described above. The final compositions of equilibrated metal and silicate at high pressure and temperature are then determined by a mass balance/element partitioning approach (*8, 14*). Water is delivered to the mantle by fully oxidized impactors, especially during the final 30-70% of accretion. The initial bulk compositions and metal-silicate equilibration pressures are refined by fitting the calculated composition of the mantle of a model Earth to that of Earth's primitive mantle or bulk silicate Earth (BSE) (*21*). As described above, best fits are obtained when bulk compositions of bodies that originated at <0.9-1.2 AU are highly reduced, bodies originating between ~1 and ~2.5 AU become increasingly oxidized with increasing heliocentric distance, and bodies from beyond 6-7 AU are fully oxidized and contain 20 wt% $H_2O$ (Fig. S1A). Refined metal-silicate equilibration pressures increase as accretion proceeds and are ~70% of the proto-Earth's core-mantle boundary pressure at the time of each impact. Equilibration temperatures are constrained to lie between the peridotite liquidus and solidus at the corresponding equilibration pressure. A list of all fitted parameters and their final values are reported below.

*Fraction of silicate mantle that equilibrates with accreted metal*: An important feature of the accretion/core formation model is that the metallic core of each impacting body sinks in a magma ocean as a mixed high-density metal-silicate plume that continuously entrains increasing amounts of silicate liquid (*13*). Metal and silicate in the plume are well mixed so that the metal equilibrates chemically with the entrained silicate liquid and there is no mixing of the metal with the rest of the mantle. The equilibrated silicate liquid constitutes only a small fraction ($X_{mantle}$) of the planet's total mantle (this contrasts strongly with the usual assumption when modelling core formation that accreted metal equilibrates chemically with the entire silicate mantle (*47*). The value of the fraction $X_{mantle}$ is determined for each impacting body from a hydrodynamic model (*13*) and, depending on the size of the impactor's core and magma ocean depth, ranges from 0.0009 to 0.008 for planetesimal impacts and 0.025 to 0.10 for giant impacts (*8*). These values are approximately consistent with the results of hydrocode calculations that also include oblique impacts (*48*). On the other hand, it has been proposed recently that planetesimal cores vaporize during high-velocity impacts, thus distributing dispersed iron over the surface of the growing Earth and enhancing the fraction of mantle that equilibrates with metal (*49*). However, to be effective, this process requires extremely high impact velocities. For example at a velocity of 17 km/s there is no vaporization and at 30-40 km/s only ~50% of the iron vaporizes (*49*). In the N-body simulation of this study, impact velocities rarely exceed 20 km/s.

*Fraction of accreted metal that equilibrates with silicate liquid*: For the metallic cores of impacting bodies to fully equilibrate with silicate liquid entrained in the metal-silicate plume described in the previous paragraph, emulsification into 1 cm size droplets is necessary (*29*). This is considered to likely occur for small planetesimal cores but it is uncertain if embryo cores fully emulsify during giant impacts (*23, 50, 51, 52*). Based on the Earth's mantle tungsten isotope anomaly, it has been estimated that 30-80% of accreted metal has equilibrated with the mantle (*52, 53*). In the combined accretion/core formation model of Rubie et al., the best least square fits to BSE composition were obtained when 80-100% of metal equilibrates with silicate (*8*). The results presented in Figs. 1, 3 and 4 are based on 100% of accreted metal equilibrating with silicate; the effect on the results of Fig. 1 when only 50% equilibrates is shown below in Fig. S2.

*Compositions of equilibrated metal and silicate liquids and the effect of accreted water*: The compositions of metallic and silicate liquids at high *P-T* conditions are determined by mass balance combined with the partitioning of major elements (Si, Fe, Ni and O) whereas trace element concentrations are based on partitioning alone (*8, 14*). Water, when delivered by oxidized CI bodies, is mixed into the mantle after each accretional event. $H_2O$ contained in the limited volume of equilibrating silicate melt oxidizes the metal through the loss of hydrogen (*8*). After each equilibration event, the metal is transferred to the core and the equilibrated silicate liquid is mixed with the rest of the mantle.

*Highly siderophile elements*: We have extended our published accretion/core formation model (*8*) in order to include the evolution of mantle Ir, Pt, Pd and Ru concentrations during Earth's accretion and differentiation. These elements include the most and the least siderophile HSEs (Ir and Pd respectively) and two elements (Pd and Ru) which are likely present in the mantle in slightly suprachondritic abundances (*32, 33*). We present results from the Grand Tack accretion simulation 4:1-0.5-8, in which a model Earth experiences a final giant impact at 113 million years (My) that increases its mass from $0.872 M_e$ to $0.997 M_e$ (*8*), consistent with Earth's Moon-forming impact ($M_e$ is Earth's current mass). The mass of material accreted after this event is 0.3% of Earth's mass ($0.003 M_e$).

*Refinement of model parameters*: The refinement of parameters in the accretion/core formation model is based primarily on fitting the calculated composition of the mantle of a model Earth to that of the bulk silicate Earth, as described previously (*8*). This is done on the basis of the elements Si, O, Fe, Ni, Co, Nb, Ta, V and Cr, with Mg, Al and Ca also being included in the bulk compositions. Five fitting parameters are refined by least squares and the resulting values for the simulation 4:1-0.5-8 studied here are as follows.

- Metal-silicate equilibration pressure = $P_f \times P_{CMB}$ with $P_f$ = 0.67 ($P_{CMB}$ is the core-mantle boundary pressure at the time of impact).

The four parameters indicated by red arrows in Fig. S1A, that define the oxygen contents of initial embryos and planetesimals, are refined in the final simulation of Fig. 4B to have the following values:

- Fraction of total Si initially contained in metal $X_{Si}^{met}(1)$ = 0.20
- Fraction of total Fe initially contained in metal $X_{Fe}^{met}(2)$ = 0.11
- Distance $\delta(1)$ = 0.95 AU
- Distance $\delta(2)$ = 2.82 AU

These values differ slightly from those reported previously (*8*) because of the incorporation of S, Pt, Pd, Ru and Ir in the present model. The distance $\delta(3)$, beyond which planetesimals contain 20 wt% water, is set at 6.8 AU in order to obtain a final mantle $H_2O$ concentration of ~1000 ppm. The incorporation of sulfide segregation in the present model has only a small effect on the parameters listed above and concentrations of the elements considered here (with the exception of S and the HSEs) because the mass fraction of segregated sulfide liquid is small. For example, refining the model without sulfide segregation results in a final mantle FeO concentration of 8.1 wt%; if the model is then re-run with the same parameter values but including FeS exsolution and segregation to the core, the final mantle FeO concentration is reduced only slightly to 7.7 wt%. Because of this small effect, we set the (unknown) Ni and oxygen contents of the segregated FeS liquid to zero. Of course, FeS segregation will have a major effect on concentrations of chalcophile trace elements, such as Cu, which are not considered here.

Two parameters that are adjusted to obtain the final mantle concentrations of the HSEs (Pt, Pd, Ru and Ir) and a mantle sulfur concentration of 200-250 ppm are:

- The effective pressure of sulfide saturation and sulphide liquid – silicate liquid equilibration in the magma ocean ($P_{eq-S} = k_S \times P_{CMB}$, with $k_S$ = 0.44).
- The time at which accretion of the late veneer starts (= 119 My in the simulation studied here).

*Metal-silicate and sulfide-silicate partitioning of Pt, Ru, Pd and Ir*: We used the experimental results of Laurenz et al. (*16*) as summarized below. The metal-silicate distribution coefficient $K_D$ is independent of oxygen fugacity and, for element M, is defined as:

$$K_D = \frac{X_M^{met}(X_{FeO}^{sil})^{n/2}}{X_{MO_{n/2}}^{sil}(X_{Fe}^{met})^{n/2}}. \quad (S1)$$

Here $X$ represents the mole fractions of M, $MO_{n/2}$, Fe and FeO in metal (met) and silicate (sil) and $n$ is the valence of M when dissolved in silicate liquid. $K_D$ for sulfide-silicate partitioning can be expressed accordingly.

*Effect of sulfur on HSE metal-silicate partitioning*: We describe this effect by:

$$\log K_D^0(metal-silicate) = a^{met} + \frac{b^{met}}{T} + \frac{c^{met}P}{T} + d^{met}\log(1-X_S) \quad (S2)$$

where $\log K_D^0(metal-silicate)$ is $\log K_D(metal-silicate)$ corrected to infinite dilution of the HSEs (*3, 16*), *a*, *b*, *c* and *d* are constants, $X_S$ is the mole fraction of S in the metal, *P* is in GPa and *T* in K. The resulting fitting parameters are listed in Table S1.

*Partitioning of HSEs between sulfide and silicate liquids*: The effects of *P* and *T* on partitioning are described by:

$$\log K_D^0(\text{sulfide}-silicate) = a^{sulf} + \frac{b^{sulf}}{T} + \frac{c^{sulf}P}{T} \quad (S3)$$

where $\log K_D^0$(sulfide-silicate) is $\log K_D$(sulfide-silicate) corrected to infinite dilution of the HSEs (*16*) and *a*, *b* and *c* are constants. The resulting fitting parameters are listed in Table S2.

*Extrapolation of partitioning data to high pressures and temperatures*: The partitioning parameters of Tables S1 and S2 have been determined experimentally at pressures ≤21 GPa. In the combined accretion/core formation models, $K_D$s are extrapolated to 80-90 GPa (metal-silicate equilibration) and ~60 GPa (sulfide-silicate equilibration) respectively. Such large extrapolations result in uncertainties that we have estimated by propagating the errors on the partitioning parameters. Experiments performed up to 100 GPa in diamond anvil cells on the metal-silicate partitioning of Ni and Co gave results that are broadly consistent with extrapolations of low-pressure data (*54, 55*

Ideally, metal-silicate partitioning experiments should be performed up to 60-100 GPa for all siderophile elements using diamond anvil cells (*55*). However, in the case of HSEs such experiments are likely to be impossible with currently available technology. This is because the heated samples in laser-heated diamond anvil cell samples are extremely small (e.g the quenched silicate region is typically around 2-3 microns across – see Fig. 2 in Fischer et al. (*55*). Analyzing ppm levels of HSEs reliably in such a small region is currently impossible, even with a nano-SIMS.

*Sulfur content at sulfide saturation (SCSS) in peridotite melt*: Laurenz et al. have determined *SCSS* for a primitive mantle composition experimentally at 2373-2773 K and 7-21 GPa (*16*). A fit to 24 experimental data points gives:

$$\ln(SCSS) = 14.2(\pm 1.18) - \frac{11032(\pm 3119)}{T} - \frac{379(\pm 82)P}{T} \quad (S4)$$

where the concentration of S is in ppm, *T* is in K and *P* is in GPa.

*Partitioning of sulfur between metal and silicate*: The metal-silicate partitioning of S has been investigated at 2073-2673 K and 2-23 GPa by Boujibar et al. (*19*). They parameterized $\log D_S^{met-sil}$ as a function of *P*, *T* and the concentrations of Fe, Ni, Si, O, P and C in the metal. We have used a modified version of their partitioning equation:

$$\log D_S^{met-sil} = \log C_S + \frac{405}{T} + \frac{136P}{T} + 32\log(1-X_{Si}) + 181\left[\log(1-X_{Si})\right]^2 + 305\left[\log(1-X_{Si})\right]^3 \quad (S5)$$
$$+ 1.13\log(1-X_{Fe}) + 10.7\log(1-X_{Ni}) - 3.72$$

where $X_{Si}$, $X_{Fe}$, and $X_{Ni}$ are the mole fractions of Si, Fe and Ni in the liquid metal. $C_S$ is the sulfide capacity of the silicate melt:

$$\log C_S = -5.704 + 3.15X_{FeO} + 2.65X_{CaO} + 0.12X_{MgO} \quad (S6)$$

where $X_{FeO}$, $X_{CaO}$ and $X_{MgO}$ are the respective mole fractions of oxide components in the silicate liquid (*19*).

We do not include the effects of C and P in the metal in Eq. S5 because these elements are not included in the accretion/core formation model. We also exclude possible effects of oxygen. According to the results of Boujibar et al., oxygen dissolved in the metal has the effect of reducing the value of $D_S^{met-sil}$ (*19*) and the reduction becomes large at high oxygen concentrations of 5-10 wt%. However, according to other studies (*56, 57*), the effect should be the opposite, with O in the metal causing $D_S^{met-sil}$ to increase. The experiments of Boujibar et al. produced very low oxygen concentrations in the metal (*19*) that are also difficult to measure; consequently, their parameterization of the effect of oxygen seems not to be reliable. Because we exclude the effects of oxygen in the metal on S partition coefficients, the values predicted for $\log D_S^{met-sil}$ by Eq. (S5) might be too low. Therefore we consider below an extreme end member case in which *all* S is partitioned into metal in each equilibration event.

In general mantle S concentrations do not decrease as a result of metal-silicate equilibration during core formation (Fig. 1). The main reason is that the metal-silicate partition coefficient of S decreases strongly with increasing Si content of the metal (in addition to other compositional factors) – see Eq. S5 and Fig. 4 in Boujibar et al. (*19*). Batches of equilibrated core-forming metal in the simulation presented here generally have Si concentrations in the range 1-12 wt%. In addition, although high pressure makes S more siderophile, high temperature has the opposite effect.

We did not consider the possible evaporative loss of S during accretional impacts because the low S-content of Earth cannot be the result of evaporation during giant impacts. The S-isotopic composition of Earth is close to that of E-chondrites (*58*). There is no indication from stable S isotopes for evaporation, as is also the case for K and Rb (*59*).

Segregation of sulfide liquid to the core

In a magma ocean in which the melt fraction is high, exsolved FeS droplets, with a stable diameter of ~0.5 cm, will sink to the base of the magma ocean with a velocity of ~0.3 m/s (*60*). Consequently, pools of FeS liquid will accumulate at the base of the magma ocean and will eventually segregate to the core through crystalline or partially molten mantle, either as sinking diapirs or by dyking (*2*). Because the density of liquid FeS is lower than that of liquid Fe-rich metal, rates of sulfide migration to the core by these mechanisms are likely to be relatively slow.

As the magma ocean starts to crystallize, FeS exsolution will be enhanced as the silicate melt fraction decreases. Provided the silicate melt fraction is high, segregation of FeS droplets can continue. However, once the silicate melt fraction has decreased to below a critical but poorly-known value (probably 30-50%), exsolved FeS liquid can no longer percolate through the partially molten aggregate and becomes trapped (61, 62). At this stage, HSEs that are added by continuing accretion remain in the mantle (late accretion stage).

Effect of incomplete equilibration of accreted metal on the evolution of HSE concentrations due to metal-silicate segregation

The results shown in Fig. 1 are based on the assumption that 100% of accreted metal equilibrates with silicate. As mentioned above, the actual value could be significantly lower if accreted metal fails to fully emulsify as impactor cores sink in the magma ocean. Figure S2 shows the results of Fig. 1 recalculated for the case that only 50% of accreted metal equilibrates. Compared with the results of Fig. 1, the concentrations of HSEs remaining in the mantle at the end of accretion as the result of metal-silicate segregation (i.e. without FeS segregation) are greatly increased. This is because significantly higher equilibration pressures are required to fit the BSE concentrations of the moderately siderophile elements ($P_{eq} = 0.77 \times P_{CMB}$, compared with $P_{eq} = 0.67 \times P_{CMB}$ when 100% of metal equilibrates). Metal-silicate partition coefficients for the HSEs are lower at the higher pressures (3) which results in the higher mantle concentrations shown in Fig. S2A. These results demonstrate that the need for FeS segregation is independent of the degree of metal equilibration.

Effect of metal-silicate partitioning on S

As discussed above, the S partitioning model used here (Eq. S5) may underestimate the value of the sulfur partition coefficient ($D_S^{met-sil}$) because no account is taken of the effect of the oxygen content of liquid metal. We therefore recalculate the results for single-stage FeS segregation shown in Fig. 4 but for the extreme case that *all* S partitions into metal in each metal-silicate equilibration event (Fig. S3).

Compared with the results obtained with the S partitioning model (Eq. S5), shown in Fig. 4, evolving mantle concentrations of S are relatively low but are still well in excess of the BSE sulfur concentration (by a factor of 2). (Note that the very limited extent of mantle equilibration, discussed above, is an important factor for this result.) Thus FeS segregation is still required in order to achieve the final BSE S concentration. The final HSE concentrations are, within error, consistent with BSE values although the fit is not as good quantitatively as that obtained with the S partitioning model (Fig. 4).

Effect of the initial distribution of sulfur in the protoplanetary disk

The initial distribution of sulfur in the protoplanetary disk, as shown in Fig. S1B, is uncertain (see above). We have therefore investigated the effect of its concentration being constant with heliocentric distance. If all starting bodies contain an S concentration corresponding to 0.075×CI, Earth's final bulk S content of 0.64 wt% is reproduced. The evolutions of concentrations of the HSEs and S in both single-stage and multi-stage sulfide segregation scenarios that result from this constant S distribution are shown in Fig. S4. The final mantle HSE concentrations are identical to those that result from the gradient model (Fig. S1B). We conclude therefore that the initial distribution of S is unimportant for modelling HSE geochemistry. The final core and mantle concentrations of S are 2.0 wt% and 140 ppm respectively – the latter value being slightly low compared with the BSE concentration of 200-250 (±40) ppm. The evolution of mantle S concentrations during accretion is significantly different from those obtained with the S gradient model (compare Figs. 4A and S4A) and, as with the gradient model, is strongly dependent on whether sulfide segregation is single- or multi-stage.

Fig. S1.

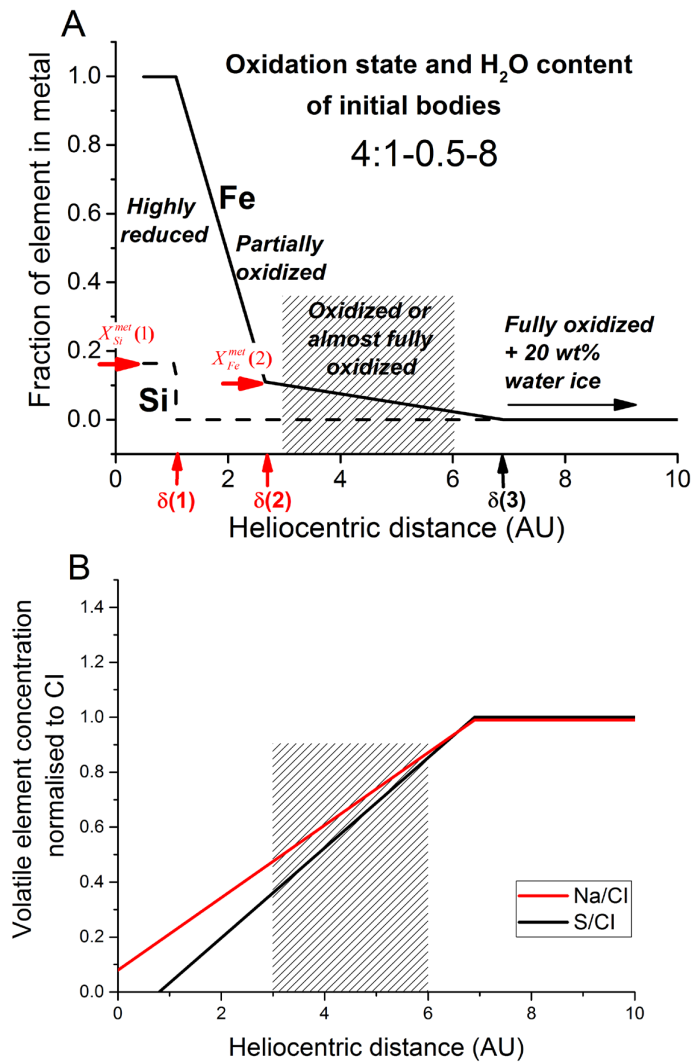

**Fig. S1.** Model of protoplanetary disk composition versus heliocentric distance that is imposed on starting embryos and planetesimals prior to the migration of Jupiter and Saturn. (A) Composition-distance model for embryos and planetesimals in the protoplanetary disk showing the variation in oxidation state and water content (*8*). Least squares fitting parameters are indicated as red arrows. (B) Compositional gradients for Na and S adopted in the present model (concentrations are normalized to CI). In the shaded region, between 3 and 6 AU, there are initially no embryos or planetesimals because all bodies in this region have been cleared by the accretion of Jupiter and Saturn.

**Fig. S2**

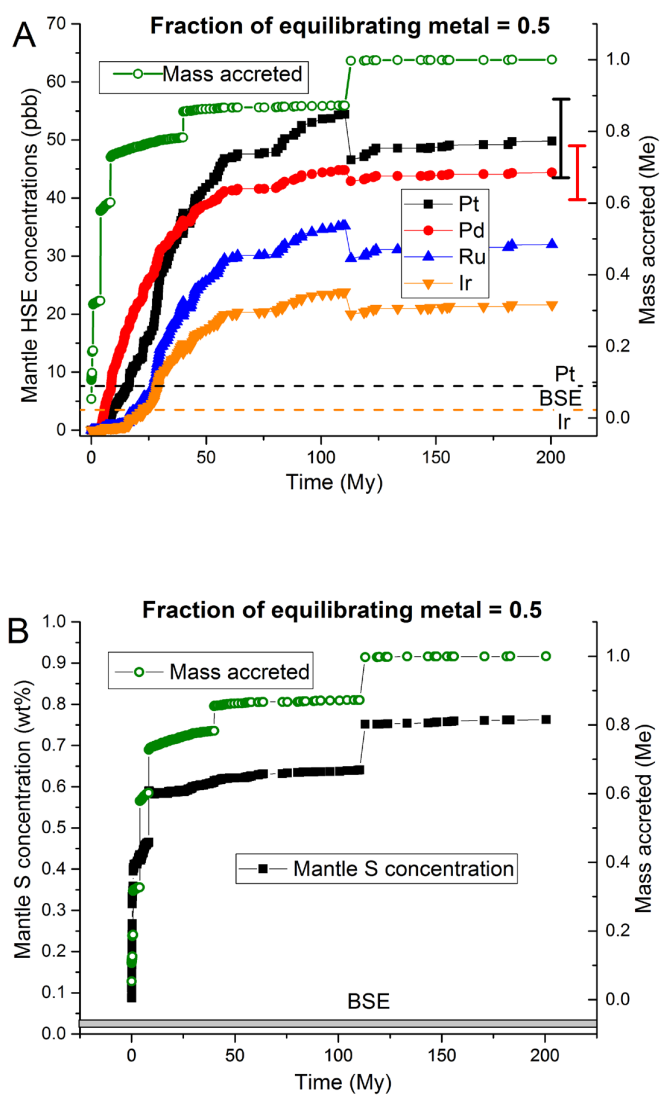

**Fig. S2**. Evolution of mass accreted and mantle abundances of (A) HSEs and (B) sulfur with time, calculated for the case that only 50% of accreted metal equilibrates with silicate. Each symbol represents an impact and "mass accreted" is the accumulated mass after each impact, normalized to Earth's current mass ($M_e$). Error bars, based on the propagation of uncertainties in the partitioning parameters, are shown for the final Pt and Pd concentrations; propagated uncertainties for Ru and Ir are ±0.7 and ±2.3 ppb respectively. See Fig. 1 for further details.

**Fig. S3**

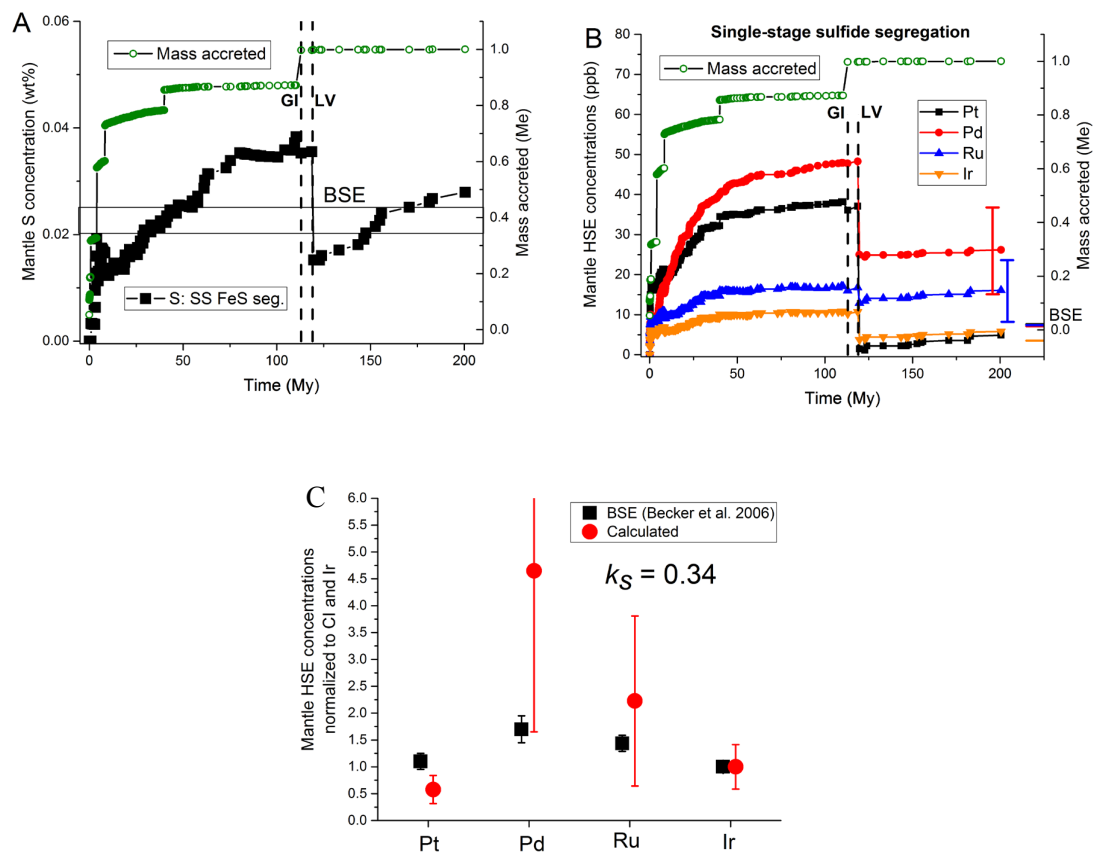

**Fig. S3.** Evolution of mantle concentrations of (A) sulfur, and (B) HSEs with time based on metal-silicate and sulfide (FeS) segregation with $k_s = 0.37$ (Eq. 1). *In contrast to Fig. 4, these results are obtained for the case that <u>all</u> S partitions into metal in each equilibration event.* The accretion history is also shown in (A, B). Results are shown for late single-stage FeS segregation at 118 My. The vertical dashed lines in (A, B) show the time of the final giant impact (GI) at 113 My and the start of late veneer accretion (LV) at 119 My. Error bars, based on the propagation of uncertainties in the partitioning parameters, are shown in (B) for the final Pd and Ru concentrations; propagated uncertainties for Pt and Ir are ±2.5 and ±1.8 ppb respectively. (C) Final calculated HSE values, normalized to Ir and CI, compared with BSE values (*32*).

**Fig. S4**

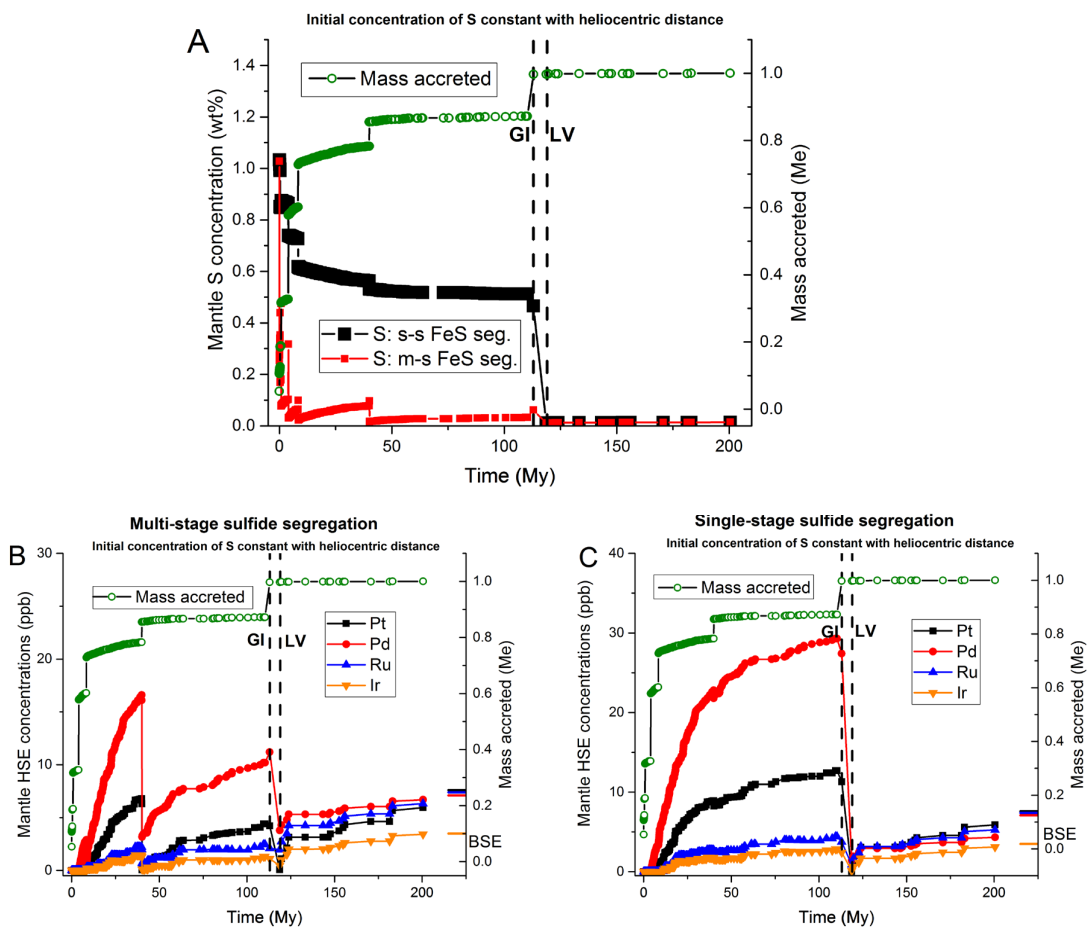

**Fig. S4.** Evolution of mantle concentrations of (A) sulfur, and (B, C) HSEs with time based on metal-silicate and sulfide (FeS) segregation with $k_s$ = 0.37 (Eq. 1). *In contrast to Fig. 4, these results are obtained for the case that all starting bodies of the protoplanetary disk have an identical bulk sulfur concentration* of 0.075×CI. For other details, see the caption of Fig. 4.

**Table S1.** Results of the regression describing the dependence of log $K_D^0(metal-silicate)$ on $P$, $T$ and $X_S$ (*17*). Values for $b^{met}$, $c^{met}$ and valence are from Mann et al. (*3*). The fits were restricted to Fe-rich metal compositions ($X_S < 0.35$). "No." is the number of measurements.

| Element | $a^{met}$ | $\pm\sigma$ | $b^{met}$ | $\pm\sigma$ | $c^{met}$ | $\pm\sigma$ | $d^{met}$ | $\pm\sigma$ | No. |
|---|---|---|---|---|---|---|---|---|---|
| Pt | -4.04 | ±0.04 | 23824 | ±211 | -17 | ±29 | 8.30 | ±3.65 | 16 |
| Pd | 0.18 | ±0.04 | 10235 | ±126 | -103 | ±25 | 9.08 | ±2.00 | 13 |
| Ru | 0.74 | ±0.05 | 12760 | ±32 | -63 | ±47 | 10.34 | ±1.29 | 12 |
| Ir | -0.67 | ±0.13 | 17526 | ±55 | -19 | ±117 | 14.50 | ±4.80 | 9 |

**Table S2.** Results of the regression describing the dependence of log $K_D^0$(sulfide-silicate) (corrected to infinite dilution) on $P$ and $T$ (*16*). "No." is the number of measurements.

| Element | $a^{sulf}$ | $\pm\sigma$ | $b^{sulf}$ | $\pm\sigma$ | $c^{sulf}$ | $\pm\sigma$ | No. |
|---|---|---|---|---|---|---|---|
| Pt | 2.04 | ±0.10 | 4306 | ±1638 | 69.4 | ±20.9 | 11 |
| Pd | 3.02 | ±0.09 | 842 | ±1653 | -19.1 | ±17.7 | 11 |
| Ru | -1.13 | ±0.12 | 12748 | ±2730 | -41.4 | ±25.2 | 7 |
| Ir | -1.16 | ±0.06 | 14073 | ±314 | -17.8 | ±13.1 | 5 |

**Additional Data Table S1 (separate file)**

Evolution of Earth's mantle composition with single-stage sulfide segregation

**Additional Data Table S2 (separate file)**

Evolution of Earth's mantle composition with multi-stage sulfide segregation

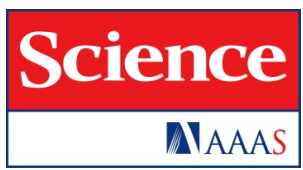

Supplementary Materials for

**Highly siderophile elements were stripped from Earth's mantle by iron sulfide segregation**

David C. Rubie, Vera Laurenz, Seth A. Jacobson, Alessandro Morbidelli, Herbert Palme, Antje K. Vogel, Daniel J Frost

correspondence to: dave.rubie@uni-bayreuth.de

**This PDF file includes:**

Supplementary Text
Figs. S1 to S4
Tables S1 to S2
Captions for databases S1 to S2

## 1. Supplementary Text

Late veneer and late accretion

These terms have been used with slightly different meanings by different authors – e.g. Walker et al. (*4*) and Jacobson et al. (*6*). Here we use "late veneer" to mean all material that was accreted to Earth after the end of core formation and sulfide segregation. We use the term "late accretion" to describe the physical process by which the late veneer was added.

Accretion/core formation model

The accretion/core formation model is identical to one of six published models (*8*) but with the addition of two volatile elements (Na and S) and four highly-siderophile elements (HSEs: Pt, Ru, Pd and Ir).

Grand Tack accretion scenario

The Grand tack scenario is the first accretion model that couples the processes of migration of the giant planets and formation of terrestrial planets. Originally proposed by Walsh et al. (*11*), this scenario is based on previous results of hydro-dynamical simulations showing that a Jupiter-mass planet alone in the protoplanetary disk would migrate inwards but the couple Jupiter-Saturn, once trapped in resonance, would migrate outwards (*40, 41, 42*). The Grand Tack scenario thus postulates that Jupiter formed first, at a distance of 3-4 AU from the Sun, and migrated inwards. Saturn, after reaching a mass close to its present one, also started to migrate towards the Sun and caught Jupiter in resonance. At this point, Jupiter and Saturn reversed the direction of migration and started to move towards the outer disk. The reversal of migration is dubbed the "tack", hence the name of the model. The embryo and planetesimal populations in the region (~3-6 AU) swept by the inwards-then-outwards migration of Jupiter are strongly depleted (Fig. S1A). The Grand Tack scenario postulates that the tack occurred when Jupiter was at about 1.5 AU from the Sun, so that the resulting depletion of mass outside of 1 AU can explain the final small mass of Mars (*43*), which could not be reproduced by earlier classical accretion models.

From the perspective of terrestrial planet formation, the Grand Tack model predicts a wide mixing of embryos and planetesimals that originally accreted in different parts of the disk. The local material, indigenous of the region ~1 AU, is mixed with material originally from the 1-3 AU region and pushed inwards by the resonances with Jupiter during the inward migration phase of the giant planet. This process of dynamical mixing also boosts the initial accretion rate of the terrestrial planets relative to the classical scenario (*44*), because it breaks the mutual dynamical isolation of the planetary embryos resulting from their growth in oligarchic fashion (*45, 46*). Finally, when the outer giant planets migrate outwards, multiple primitive and water-rich planetesimals from the giant planet zone are scattered towards the terrestrial planet region, thus contributing significantly to Earth's water budget (*9*).

The Grand Tack scenario is also successful in explaining the orbital distribution of the terrestrial planets (*10*), such as their low angular momentum deficit (which is a measure of the small deviations of the terrestrial planet orbits from perfect co-planar circles). Provided some assumptions hold true on the mass ratio between embryos and planetesimals, it also predicts that the growth of the Earth, although initially very rapid, can be protracted for ~100 My; in this case the amount of material accreted after the last giant impact is small (less than 1% of the Earth-mass), consistent with the amount of "late veneer" inferred from mantle HSE abundances (*5, 6, 33*).

From the compositional point of view, given reasonable initial chemical gradients in the disk, as outlined below, the Grand Tack scenario is consistent with the chemistry of the major non-volatile lithophile and weakly-siderophile elements in the mantle and the oxidation state of the Earth (*8*).

Structure and chemistry of the protoplanetary disk

The Grand Tack simulation selected for this study is 4:1-0.5-8 because it produced a model Earth close to 1 AU with a final mass close to one Earth mass ($M_e$). In addition, the final giant impact occurs at 113 My and the mass of material accreted following this impact is about $0.003 M_e$. In the simulation number "4:1-0.5-8", "4:1" indicates the ratio of the total masses in the embryo and planetesimal populations respectively, and "0.5" signifies that the initial embryo mass is 0.5 × mass of Mars. Based on these parameters, ten simulations were run with very slight variations in the initial orbital characteristics of the starting bodies: the last term ("8") is the run number within the set of 10 simulations. The protoplanetary disk consists of 87 embryos, distributed between 0.7-3.0 AU, and 2836 planetesimals, each of mass $3.9 \times 10^{-4} M_e$, initially distributed between 0.7-3.0 AU and 6.0-9.5 AU (the region between 3.0 and 6.0 AU is cleared of planetesimals by the accretion of Jupiter and Saturn).

The bulk chemistry of each of the embryos and planetesimals is specified <u>prior to the migration of Jupiter and Saturn</u> by assuming Solar System bulk composition ratios (CI chondritic) for all non-volatile elements (Mg, Si, Ni, Co, Cr and Pd) but with refractory element concentrations (Ca, Al, Nb, Ta, Pt, Ru and Ir) that are enhanced by 22% (11% for V) relative to CI (normalized to Mg), as justified previously (*14*). Oxygen and water contents are two critical compositional variables. The composition of a model Earth's mantle, calculated from our accretion/core formation model, is fit to that of the bulk silicate Earth (BSE). The exploration of a broad parameter space results in the optimal composition-distance model for primitive bodies shown in Fig. S1A (*8*). Here compositions close to the Sun (<1.1 AU) are highly reduced with 99.9% of Fe being present as metal and with ~18% of total Si initially dissolved in the metal. At heliocentric distances >1.1 AU, compositions become increasing oxidized and beyond 6.9 AU they are fully oxidized and contain 20 wt% $H_2O$. Four least-squares fitting parameters are indicated in Fig. S1A as red arrows and consist of two compositional parameters and two distance parameters. The distance $\delta(3)$ was fixed to be 6.9 AU by requiring a final mantle $H_2O$ content of ~1000 ppm.

Using a similar approach, we include the volatile element S (and also for comparison the lithophile/volatile element Na) in the current model by imposing compositional gradients on starting bodies (Fig. S1B). These gradients reflect high volatility at low heliocentric distances because of high temperatures and an increased tendency to condense as temperatures decrease with increasing heliocentric distance in the nebula (*18*). The Na gradient is adjusted in order to obtain a final mantle (BSE) concentration of 2600 ppm (*21*) assuming that Na does not partition into the core. In the case of S, the aim is to obtain the bulk Earth's S content which results when bodies at ≤0.8 AU contain zero

S. The steeper gradient for S compared to Na is qualitatively consistent with the lower 50% condensation temperatures of S at $10^{-4}$ bar of 655 K compared to 953 K for Na. (*17*).

The sulfur gradient of Fig. S1B is likely to be an oversimplification of reality. For example, it implies that highly reduced compositions from ≤1 AU should contain no S – whereas enstatite chondrites, which are highly reduced, contain up to 5-6 wt% S. Such discrepancies may possibly be a consequence of the evolution of protoplanetary disk chemistry with time. We show below that provided there is sufficient S to provide Earth with its bulk S content (0.64 wt%), our main conclusions are independent of its distribution in the protoplanetary disk.

Core formation model

We modelled the evolution of some moderately siderophile non-volatile element abundances in Earth's mantle and core by integrating the dynamical process of planetary accretion with the chemistry of core-mantle differentiation (*8*). Planets grow through collisions with embryos and planetesimals which (apart from fully-oxidized C-type bodies) are assumed to be differentiated into core and mantle prior to accretion. Each impact delivers mass and energy to a growing planet, with the energy resulting in melting, magma ocean formation and an episode of core formation. The initial bulk compositions of all bodies are defined in terms of oxygen and $H_2O$ contents, as described above. The final compositions of equilibrated metal and silicate at high pressure and temperature are then determined by a mass balance/element partitioning approach (*8, 14*). Water is delivered to the mantle by fully oxidized impactors, especially during the final 30-70% of accretion. The initial bulk compositions and metal-silicate equilibration pressures are refined by fitting the calculated composition of the mantle of a model Earth to that of Earth's primitive mantle or bulk silicate Earth (BSE) (*21*). As described above, best fits are obtained when bulk compositions of bodies that originated at <0.9-1.2 AU are highly reduced, bodies originating between ~1 and ~2.5 AU become increasingly oxidized with increasing heliocentric distance, and bodies from beyond 6-7 AU are fully oxidized and contain 20 wt% $H_2O$ (Fig. S1A). Refined metal-silicate equilibration pressures increase as accretion proceeds and are ~70% of the proto-Earth's core-mantle boundary pressure at the time of each impact. Equilibration temperatures are constrained to lie between the peridotite liquidus and solidus at the corresponding equilibration pressure. A list of all fitted parameters and their final values are reported below.

*Fraction of silicate mantle that equilibrates with accreted metal*: An important feature of the accretion/core formation model is that the metallic core of each impacting body sinks in a magma ocean as a mixed high-density metal-silicate plume that continuously entrains increasing amounts of silicate liquid (*13*). Metal and silicate in the plume are well mixed so that the metal equilibrates chemically with the entrained silicate liquid and there is no mixing of the metal with the rest of the mantle. The equilibrated silicate liquid constitutes only a small fraction ($X_{mantle}$) of the planet's total mantle (this contrasts strongly with the usual assumption when modelling core formation that accreted metal equilibrates chemically with the entire silicate mantle (*47*). The value of the fraction $X_{mantle}$ is determined for each impacting body from a hydrodynamic model (*13*) and, depending on the size of the impactor's core and magma ocean depth, ranges from 0.0009 to 0.008 for planetesimal impacts and 0.025 to 0.10 for giant impacts (*8*). These values are approximately consistent with the results of hydrocode calculations that also include oblique impacts (*48*). On the other hand, it has been proposed recently that planetesimal cores vaporize during high-velocity impacts, thus distributing dispersed iron over the surface of the growing Earth and enhancing the fraction of mantle that equilibrates with metal (*49*). However, to be effective, this process requires extremely high impact velocities. For example at a velocity of 17 km/s there is no vaporization and at 30-40 km/s only ~50% of the iron vaporizes (*49*). In the N-body simulation of this study, impact velocities rarely exceed 20 km/s.

*Fraction of accreted metal that equilibrates with silicate liquid*: For the metallic cores of impacting bodies to fully equilibrate with silicate liquid entrained in the metal-silicate plume described in the previous paragraph, emulsification into 1 cm size droplets is necessary (*29*). This is considered to likely occur for small planetesimal cores but it is uncertain if embryo cores fully emulsify during giant impacts (*23, 50, 51, 52*). Based on the Earth's mantle tungsten isotope anomaly, it has been estimated that 30-80% of accreted metal has equilibrated with the mantle (*52, 53*). In the combined accretion/core formation model of Rubie et al., the best least square fits to BSE composition were obtained when 80-100% of metal equilibrates with silicate (*8*). The results presented in Figs. 1, 3 and 4 are based on 100% of accreted metal equilibrating with silicate; the effect on the results of Fig. 1 when only 50% equilibrates is shown below in Fig. S2.

*Compositions of equilibrated metal and silicate liquids and the effect of accreted water*: The compositions of metallic and silicate liquids at high *P-T* conditions are determined by mass balance combined with the partitioning of major elements (Si, Fe, Ni and O) whereas trace element concentrations are based on partitioning alone (*8, 14*). Water, when delivered by oxidized CI bodies, is mixed into the mantle after each accretional event. $H_2O$ contained in the limited volume of equilibrating silicate melt oxidizes the metal through the loss of hydrogen (*8*). After each equilibration event, the metal is transferred to the core and the equilibrated silicate liquid is mixed with the rest of the mantle.

*Highly siderophile elements*: We have extended our published accretion/core formation model (*8*) in order to include the evolution of mantle Ir, Pt, Pd and Ru concentrations during Earth's accretion and differentiation. These elements include the most and the least siderophile HSEs (Ir and Pd respectively) and two elements (Pd and Ru) which are likely present in the mantle in slightly suprachondritic abundances (*32, 33*). We present results from the Grand Tack accretion simulation 4:1-0.5-8, in which a model Earth experiences a final giant impact at 113 million years (My) that increases its mass from $0.872 M_e$ to $0.997 M_e$ (*8*), consistent with Earth's Moon-forming impact ($M_e$ is Earth's current mass). The mass of material accreted after this event is 0.3% of Earth's mass ($0.003 M_e$).

*Refinement of model parameters*: The refinement of parameters in the accretion/core formation model is based primarily on fitting the calculated composition of the mantle of a model Earth to that of the bulk silicate Earth, as described previously (*8*). This is done on the basis of the elements Si, O, Fe, Ni, Co, Nb, Ta, V and Cr, with Mg, Al and Ca also being included in the bulk compositions. Five fitting parameters are refined by least squares and the resulting values for the simulation 4:1-0.5-8 studied here are as follows.

- Metal-silicate equilibration pressure = $P_f \times P_{CMB}$ with $P_f = 0.67$ ($P_{CMB}$ is the core-mantle boundary pressure at the time of impact).

The four parameters indicated by red arrows in Fig. S1A, that define the oxygen contents of initial embryos and planetesimals, are refined in the final simulation of Fig. 4B to have the following values:

- Fraction of total Si initially contained in metal $X_{Si}^{met}(1) = 0.20$
- Fraction of total Fe initially contained in metal $X_{Fe}^{met}(2) = 0.11$
- Distance $\delta(1) = 0.95$ AU
- Distance $\delta(2) = 2.82$ AU

These values differ slightly from those reported previously (*8*) because of the incorporation of S, Pt, Pd, Ru and Ir in the present model. The distance $\delta(3)$, beyond which planetesimals contain 20 wt% water, is set at 6.8 AU in order to obtain a final mantle $H_2O$ concentration of ~1000 ppm. The incorporation of sulfide segregation in the present model has only a small effect on the parameters listed above and concentrations of the elements considered here (with the exception of S and the HSEs) because the mass fraction of segregated sulfide liquid is small. For example, refining the model without sulfide segregation results in a final mantle FeO concentration of 8.1 wt%; if the model is then re-run with the same parameter values but including FeS exsolution and segregation to the core, the final mantle FeO concentration is reduced only slightly to 7.7 wt%. Because of this small effect, we set the (unknown) Ni and oxygen contents of the segregated FeS liquid to zero. Of course, FeS segregation will have a major effect on concentrations of chalcophile trace elements, such as Cu, which are not considered here.

Two parameters that are adjusted to obtain the final mantle concentrations of the HSEs (Pt, Pd, Ru and Ir) and a mantle sulfur concentration of 200-250 ppm are:

- The effective pressure of sulfide saturation and sulphide liquid – silicate liquid equilibration in the magma ocean ($P_{eq-S} = k_S \times P_{CMB}$, with $k_S = 0.44$).
- The time at which accretion of the late veneer starts (= 119 My in the simulation studied here).

*Metal-silicate and sulfide-silicate partitioning of Pt, Ru, Pd and Ir*: We used the experimental results of Laurenz et al. (*16*) as summarized below. The metal-silicate distribution coefficient $K_D$ is independent of oxygen fugacity and, for element M, is defined as:

$$K_D = \frac{X_M^{met}(X_{FeO}^{sil})^{n/2}}{X_{MO_{n/2}}^{sil}(X_{Fe}^{met})^{n/2}}. \qquad (S1)$$

Here $X$ represents the mole fractions of M, $MO_{n/2}$, Fe and FeO in metal (met) and silicate (sil) and $n$ is the valence of M when dissolved in silicate liquid. $K_D$ for sulfide-silicate partitioning can be expressed accordingly.

*Effect of sulfur on HSE metal-silicate partitioning*: We describe this effect by:

$$\log K_D^0(metal-silicate) = a^{met} + \frac{b^{met}}{T} + \frac{c^{met}P}{T} + d^{met}\log(1-X_S) \quad (S2)$$

where $\log K_D^0(metal-silicate)$ is $\log K_D(metal-silicate)$ corrected to infinite dilution of the HSEs (*3, 16*), *a*, *b*, *c* and *d* are constants, $X_S$ is the mole fraction of S in the metal, *P* is in GPa and *T* in K. The resulting fitting parameters are listed in Table S1.

*Partitioning of HSEs between sulfide and silicate liquids*: The effects of *P* and *T* on partitioning are described by:

$$\log K_D^0(\text{sulfide}-silicate) = a^{sulf} + \frac{b^{sulf}}{T} + \frac{c^{sulf}P}{T} \quad (S3)$$

where $\log K_D^0(\text{sulfide-silicate})$ is $\log K_D(\text{sulfide-silicate})$ corrected to infinite dilution of the HSEs (*16*) and *a*, *b* and *c* are constants. The resulting fitting parameters are listed in Table S2.

*Extrapolation of partitioning data to high pressures and temperatures*: The partitioning parameters of Tables S1 and S2 have been determined experimentally at pressures ≤21 GPa. In the combined accretion/core formation models, $K_D$s are extrapolated to 80-90 GPa (metal-silicate equilibration) and ~60 GPa (sulfide-silicate equilibration) respectively. Such large extrapolations result in uncertainties that we have estimated by propagating the errors on the partitioning parameters. Experiments performed up to 100 GPa in diamond anvil cells on the metal-silicate partitioning of Ni and Co gave results that are broadly consistent with extrapolations of low-pressure data (*54 ,55*

Ideally, metal-silicate partitioning experiments should be performed up to 60-100 GPa for all siderophile elements using diamond anvil cells (*55*). However, in the case of HSEs such experiments are likely to be impossible with currently available technology. This is because the heated samples in laser-heated diamond anvil cell samples are extremely small (e.g the quenched silicate region is typically around 2-3 microns across – see Fig. 2 in Fischer et al. (*55*). Analyzing ppm levels of HSEs reliably in such a small region is currently impossible, even with a nano-SIMS.

*Sulfur content at sulfide saturation (SCSS) in peridotite melt*: Laurenz et al. have determined *SCSS* for a primitive mantle composition experimentally at 2373-2773 K and 7-21 GPa (*16*). A fit to 24 experimental data points gives:

$$\ln(SCSS) = 14.2(\pm 1.18) - \frac{11032(\pm 3119)}{T} - \frac{379(\pm 82)P}{T} \quad (S4)$$

where the concentration of S is in ppm, *T* is in K and *P* is in GPa.

*Partitioning of sulfur between metal and silicate*: The metal-silicate partitioning of S has been investigated at 2073-2673 K and 2-23 GPa by Boujibar et al. (*19*). They parameterized $\log D_S^{met-sil}$ as a function of *P*, *T* and the concentrations of Fe, Ni, Si, O, P and C in the metal. We have used a modified version of their partitioning equation:

$$\log D_S^{met-sil} = \log C_S + \frac{405}{T} + \frac{136P}{T} + 32\log(1-X_{Si}) + 181[\log(1-X_{Si})]^2 + 305[\log(1-X_{Si})]^3 \quad (S5)$$
$$+ 1.13\log(1-X_{Fe}) + 10.7\log(1-X_{Ni}) - 3.72$$

where $X_{Si}$, $X_{Fe}$, and $X_{Ni}$ are the mole fractions of Si, Fe and Ni in the liquid metal. $C_S$ is the sulfide capacity of the silicate melt:

$$\log C_S = -5.704 + 3.15 X_{FeO} + 2.65 X_{CaO} + 0.12 X_{MgO} \quad (S6)$$

where $X_{FeO}$, $X_{CaO}$ and $X_{MgO}$ are the respective mole fractions of oxide components in the silicate liquid (*19*).

We do not include the effects of C and P in the metal in Eq. S5 because these elements are not included in the accretion/core formation model. We also exclude possible effects of oxygen. According to the results of Boujibar et al., oxygen dissolved in the metal has the effect of reducing the value of $D_S^{met-sil}$ (*19*) and the reduction becomes large at high oxygen concentrations of 5-10 wt%. However, according to other studies (*56, 57*), the effect should be the opposite, with O in the metal causing $D_S^{met-sil}$ to increase. The experiments of Boujibar et al. produced very low oxygen concentrations in the metal (*19*) that are also difficult to measure; consequently, their parameterization of the effect of oxygen seems not to be reliable. Because we exclude the effects of oxygen in the metal on S partition coefficients, the values predicted for $\log D_S^{met-sil}$ by Eq. (S5) might be too low. Therefore we consider below an extreme end member case in which *all* S is partitioned into metal in each equilibration event.

In general mantle S concentrations do not decrease as a result of metal-silicate equilibration during core formation (Fig. 1). The main reason is that the metal-silicate partition coefficient of S decreases strongly with increasing Si content of the metal (in addition to other compositional factors) – see Eq. S5 and Fig. 4 in Boujibar et al. (*19*). Batches of equilibrated core-forming metal in the simulation presented here generally have Si concentrations in the range 1-12 wt%. In addition, although high pressure makes S more siderophile, high temperature has the opposite effect.

We did not consider the possible evaporative loss of S during accretional impacts because the low S-content of Earth cannot be the result of evaporation during giant impacts. The S-isotopic composition of Earth is close to that of E-chondrites (*58*). There is no indication from stable S isotopes for evaporation, as is also the case for K and Rb (*59*).

Segregation of sulfide liquid to the core

In a magma ocean in which the melt fraction is high, exsolved FeS droplets, with a stable diameter of ~0.5 cm, will sink to the base of the magma ocean with a velocity of ~0.3 m/s (*60*). Consequently, pools of FeS liquid will accumulate at the base of the magma ocean and will eventually segregate to the core through crystalline or partially molten mantle, either as sinking diapirs or by dyking (*2*). Because the density of liquid FeS is lower than that of liquid Fe-rich metal, rates of sulfide migration to the core by these mechanisms are likely to be relatively slow.

As the magma ocean starts to crystallize, FeS exsolution will be enhanced as the silicate melt fraction decreases. Provided the silicate melt fraction is high, segregation of FeS droplets can continue. However, once the silicate melt fraction has decreased to below a critical but poorly-known value (probably 30-50%), exsolved FeS liquid can no longer percolate through the partially molten aggregate and becomes trapped (61, 62). At this stage, HSEs that are added by continuing accretion remain in the mantle (late accretion stage).

Effect of incomplete equilibration of accreted metal on the evolution of HSE concentrations due to metal-silicate segregation

The results shown in Fig. 1 are based on the assumption that 100% of accreted metal equilibrates with silicate. As mentioned above, the actual value could be significantly lower if accreted metal fails to fully emulsify as impactor cores sink in the magma ocean. Figure S2 shows the results of Fig. 1 recalculated for the case that only 50% of accreted metal equilibrates. Compared with the results of Fig. 1, the concentrations of HSEs remaining in the mantle at the end of accretion as the result of metal-silicate segregation (i.e. without FeS segregation) are greatly increased. This is because significantly higher equilibration pressures are required to fit the BSE concentrations of the moderately siderophile elements ($P_{eq} = 0.77 \times P_{CMB}$, compared with $P_{eq} = 0.67 \times P_{CMB}$ when 100% of metal equilibrates). Metal-silicate partition coefficients for the HSEs are lower at the higher pressures (*3*) which results in the higher mantle concentrations shown in Fig. S2A. These results demonstrate that the need for FeS segregation is independent of the degree of metal equilibration.

Effect of metal-silicate partitioning on S

As discussed above, the S partitioning model used here (Eq. S5) may underestimate the value of the sulfur partition coefficient ($D_S^{met-sil}$) because no account is taken of the effect of the oxygen content of liquid metal. We therefore recalculate the results for single-stage FeS segregation shown in Fig. 4 but for the extreme case that *all* S partitions into metal in each metal-silicate equilibration event (Fig. S3).

Compared with the results obtained with the S partitioning model (Eq. S5), shown in Fig. 4, evolving mantle concentrations of S are relatively low but are still well in excess of the BSE sulfur concentration (by a factor of 2). (Note that the very limited extent of mantle equilibration, discussed above, is an important factor for this result.) Thus FeS segregation is still required in order to achieve the final BSE S concentration. The final HSE concentrations are, within error, consistent with BSE values although the fit is not as good quantitatively as that obtained with the S partitioning model (Fig. 4).

Effect of the initial distribution of sulfur in the protoplanetary disk

The initial distribution of sulfur in the protoplanetary disk, as shown in Fig. S1B, is uncertain (see above). We have therefore investigated the effect of its concentration being constant with heliocentric distance. If all starting bodies contain an S concentration corresponding to 0.075×CI, Earth's final bulk S content of 0.64 wt% is reproduced. The evolutions of concentrations of the HSEs and S in both single-stage and multi-stage sulfide segregation scenarios that result from this constant S distribution are shown in Fig. S4. The final mantle HSE concentrations are identical to those that result from the gradient model (Fig. S1B). We conclude therefore that the initial distribution of S is unimportant for modelling HSE geochemistry. The final core and mantle concentrations of S are 2.0 wt% and 140 ppm respectively – the latter value being slightly low compared with the BSE concentration of 200-250 (±40) ppm. The evolution of mantle S concentrations during accretion is significantly different from those obtained with the S gradient model (compare Figs. 4A and S4A) and, as with the gradient model, is strongly dependent on whether sulfide segregation is single- or multi-stage.

Fig. S1.

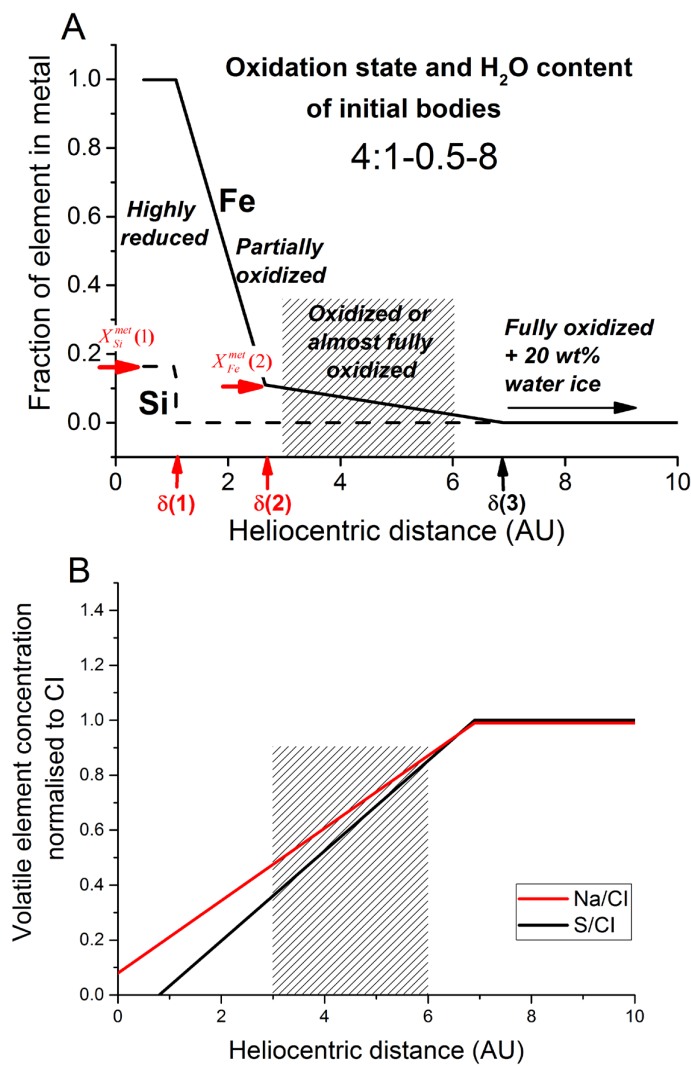

**Fig. S1.** Model of protoplanetary disk composition versus heliocentric distance that is imposed on starting embryos and planetesimals prior to the migration of Jupiter and Saturn. (A) Composition-distance model for embryos and planetesimals in the protoplanetary disk showing the variation in oxidation state and water content (*8*). Least squares fitting parameters are indicated as red arrows. (B) Compositional gradients for Na and S adopted in the present model (concentrations are normalized to CI). In the shaded region, between 3 and 6 AU, there are initially no embryos or planetesimals because all bodies in this region have been cleared by the accretion of Jupiter and Saturn.

**Fig. S2**

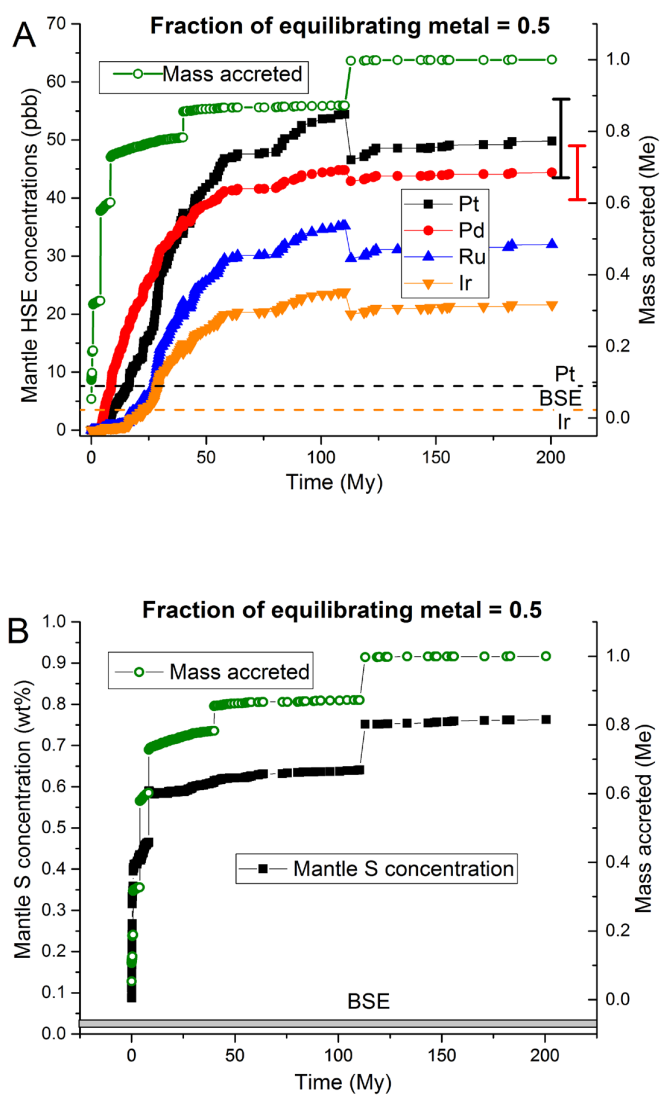

**Fig. S2**. Evolution of mass accreted and mantle abundances of (A) HSEs and (B) sulfur with time, calculated for the case that only 50% of accreted metal equilibrates with silicate. Each symbol represents an impact and "mass accreted" is the accumulated mass after each impact, normalized to Earth's current mass ($M_e$). Error bars, based on the propagation of uncertainties in the partitioning parameters, are shown for the final Pt and Pd concentrations; propagated uncertainties for Ru and Ir are ±0.7 and ±2.3 ppb respectively. See Fig. 1 for further details.

**Fig. S3**

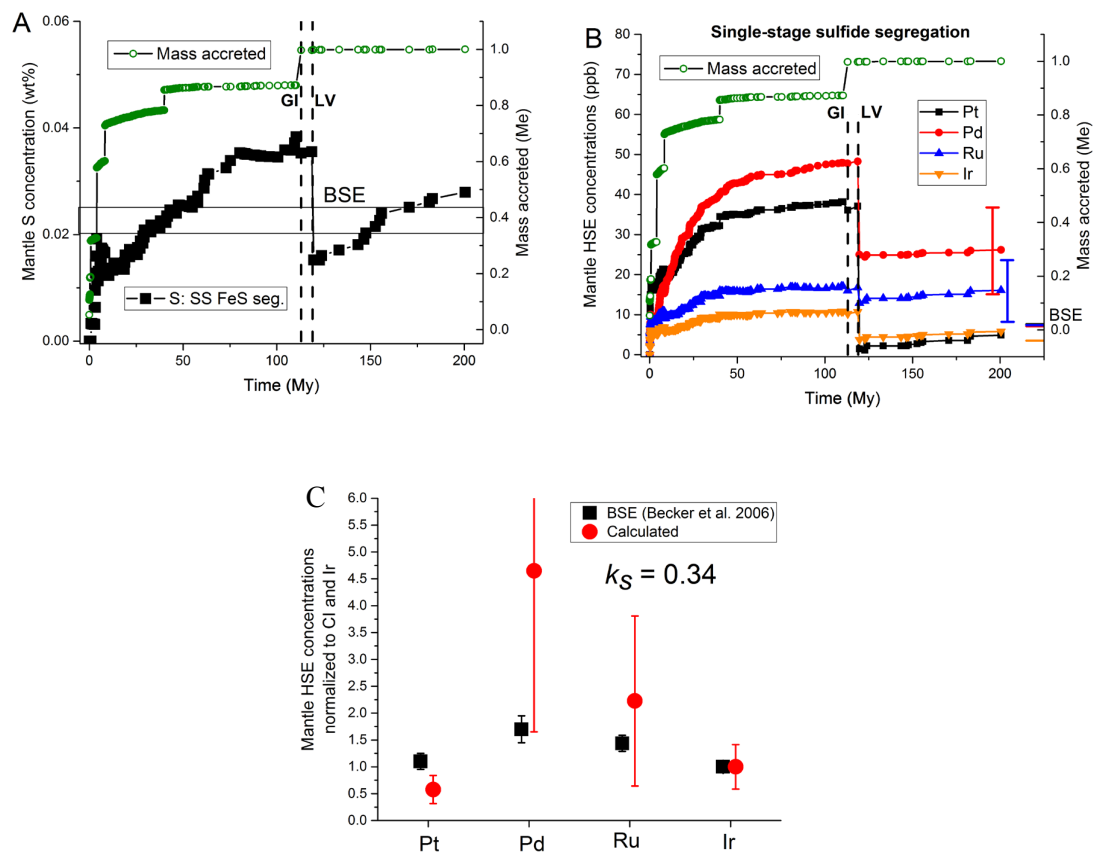

**Fig. S3.** Evolution of mantle concentrations of (A) sulfur, and (B) HSEs with time based on metal-silicate and sulfide (FeS) segregation with $k_s = 0.37$ (Eq. 1). *In contrast to Fig. 4, these results are obtained for the case that <u>all</u> S partitions into metal in each equilibration event.* The accretion history is also shown in (A, B). Results are shown for late single-stage FeS segregation at 118 My. The vertical dashed lines in (A, B) show the time of the final giant impact (GI) at 113 My and the start of late veneer accretion (LV) at 119 My. Error bars, based on the propagation of uncertainties in the partitioning parameters, are shown in (B) for the final Pd and Ru concentrations; propagated uncertainties for Pt and Ir are ±2.5 and ±1.8 ppb respectively. (C) Final calculated HSE values, normalized to Ir and CI, compared with BSE values (*32*).

**Fig. S4**

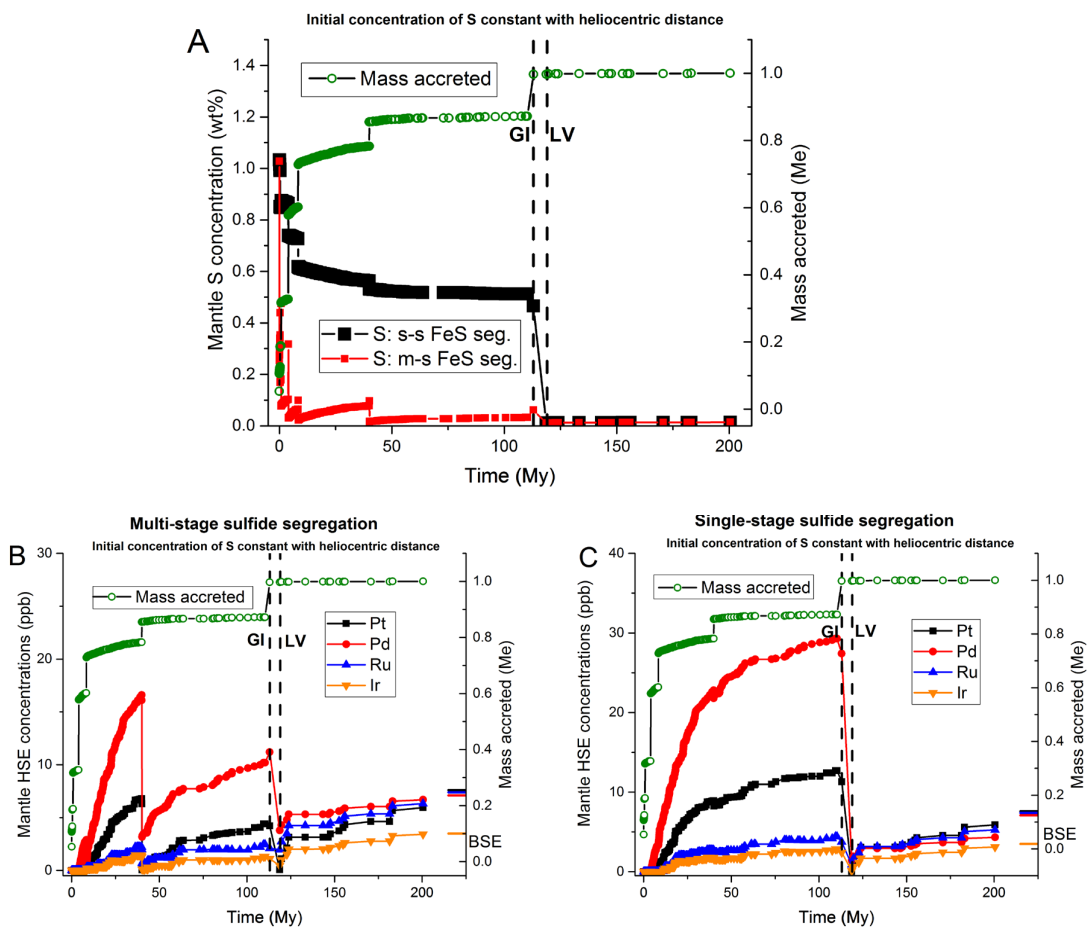

**Fig. S4.** Evolution of mantle concentrations of (A) sulfur, and (B, C) HSEs with time based on metal-silicate and sulfide (FeS) segregation with $k_s = 0.37$ (Eq. 1). *In contrast to Fig. 4, these results are obtained for the case that all starting bodies of the protoplanetary disk have an identical bulk sulfur concentration* of 0.075×CI. For other details, see the caption of Fig. 4.

**Table S1.** Results of the regression describing the dependence of log $K_D^0(metal-silicate)$ on $P$, $T$ and $X_S$ (*17*). Values for $b^{met}$, $c^{met}$ and valence are from Mann et al. (*3*). The fits were restricted to Fe-rich metal compositions ($X_S < 0.35$). "No." is the number of measurements.

| Element | $a^{met}$ | ±σ | $b^{met}$ | ±σ | $c^{met}$ | ±σ | $d^{met}$ | ±σ | No. |
|---|---|---|---|---|---|---|---|---|---|
| Pt | -4.04 | ±0.04 | 23824 | ±211 | -17 | ±29 | 8.30 | ±3.65 | 16 |
| Pd | 0.18 | ±0.04 | 10235 | ±126 | -103 | ±25 | 9.08 | ±2.00 | 13 |
| Ru | 0.74 | ±0.05 | 12760 | ±32 | -63 | ±47 | 10.34 | ±1.29 | 12 |
| Ir | -0.67 | ±0.13 | 17526 | ±55 | -19 | ±117 | 14.50 | ±4.80 | 9 |

**Table S2.** Results of the regression describing the dependence of log $K_D^0$(sulfide-silicate) (corrected to infinite dilution) on $P$ and $T$ (*16*). "No." is the number of measurements.

| Element | $a^{sulf}$ | $\pm\sigma$ | $b^{sulf}$ | $\pm\sigma$ | $c^{sulf}$ | $\pm\sigma$ | No. |
|---|---|---|---|---|---|---|---|
| Pt | 2.04 | ±0.10 | 4306 | ±1638 | 69.4 | ±20.9 | 11 |
| Pd | 3.02 | ±0.09 | 842 | ±1653 | -19.1 | ±17.7 | 11 |
| Ru | -1.13 | ±0.12 | 12748 | ±2730 | -41.4 | ±25.2 | 7 |
| Ir | -1.16 | ±0.06 | 14073 | ±314 | -17.8 | ±13.1 | 5 |

**Additional Data Table S1 (separate file)**

Evolution of Earth's mantle composition with single-stage sulfide segregation

**Additional Data Table S2 (separate file)**

Evolution of Earth's mantle composition with multi-stage sulfide segregation